\documentclass[twocolumn,aps,superscriptaddress,multicol,amsmath,amssymb]{revtex4-2}
\usepackage{amssymb}
\usepackage{amsmath}
\usepackage{graphicx}
\usepackage{sidecap}
\usepackage{epstopdf}
\usepackage{bm}% 
\usepackage{gensymb}
\usepackage{soul}
\usepackage{sidecap}
\usepackage{bbm}
\usepackage[normalem]{ulem}
\UseRawInputEncoding
\newcommand{\bra}[1]{\ensuremath{\langle #1 |}}
\newcommand{\ket}[1]{\ensuremath{| #1 \rangle}}

\newcommand{\be}{\begin{equation}}
	\newcommand{\ee}{\end{equation}}
\newcommand{\bk}{{{\bf{k}}}}
\newcommand{\bq}{{{\bf{q}}}}
\newcommand{\bK}{{{\bf{K}}}}

\newcommand{\br}{{{\bf{r}}}}

\newcommand{\bg}{{{\bf{g}}}}

\newcommand{\bM}{{{\bf{M}}}}

\newcommand{\bR}{{{\bf{R}}}}

\newcommand{\bea}{\begin{eqnarray}}
	\newcommand{\eea}{\end{eqnarray}}

\newcommand{\bS}{{\bf S}}

\newcommand{\bd}{\begin{displaymath}}
	\newcommand{\ed}{\end{displaymath}}
\newcommand{\ba}{\begin{array}}
	\newcommand{\ea}{\end{array}}
\newcommand{\bi}{\begin{itemize}}
	\newcommand{\ei}{\end{itemize}}
\newcommand{\bc}{\begin{center}}
	\newcommand{\ec}{\end{center}}
\newcommand{\bfl}{\begin{flushleft}}
	\newcommand{\efl}{\end{flushleft}}
\newcommand{\bfr}{\begin{flushright}}
	\newcommand{\efr}{\end{flushright}}

\newcommand{\mi}{\rm i}

\newcommand{\bl}{\begin{aligned}}
	\newcommand{\el}{\end{aligned}}

\def\ket#1{\left\vert #1 \right\rangle}
\def\br{{\bf r}}\def\bR{{\bf R}}
\def\bk{{\bf k}} \def\bK{{\bf K}}\def\bq{{\bf q}}  
\def\bg{{\bf g}}  \def\bd{{\bf d}}  \def\bS{{\bf S}}
 \def\bS{{\bf S}}

\def\6{\partial}

\def\bra{\langle}
\def\ket{\rangle}
\def\={\!\!\!&=&\!\!\!}
\def\+{\!\!\!&&\!\!\!+~}
\def\-{\!\!\!&&\!\!\!-~}

\newcommand\redout{\bgroup\markoverwith{\textcolor{red}{\rule[.5ex]{2pt}{0.4pt}}}\ULon}

\usepackage[colorlinks=true,citecolor=blue]{hyperref}
\hypersetup{colorlinks=true,citecolor=blue,linkcolor=red,urlcolor=blue}

\newcommand{\orcid}[1]{\href{https://orcid.org/#1}{\includegraphics[width=8pt]{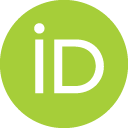}}}

\newcommand\bluesout{\bgroup\markoverwith{\textcolor{blue}{\rule[0.5ex]{2pt}{0.4pt}}}\ULon}
\begin{document}
%	\title{Pairing in the two-dimensional triangular Rashba-Hubbard model:\\ Applications to twisted bilayer transition metal dichalcogenides}

\title{Proximity-driven ferromagnetism and superconductivity in the triangular Rashba-Hubbard model} %: %Application to twisted bilayer of transition metal dichalcogenides}
	\author{Mehdi Biderang \orcid{0000-0002-6666-1659}} 
	\email{mehdi.biderang@umanitoba.ca}
	\affiliation{Department of Physics and Astronomy, University of Manitoba, Winnipeg R3T 2N2, Canada}
	\affiliation{Manitoba Quantum Institute, University of Manitoba, Winnipeg R3T 2N2, Canada}
	\author{Mohammad-Hossein Zare \orcid{0000-0003-4638-7987}}
	\email{zare@qut.ac.ir}
	\affiliation{Department of Physics, Qom University of Technology, Qom 37181-46645, Iran}
	\author{Jesko Sirker}
	\email{sirker@physics.umanitoba.ca}
	\affiliation{Department of Physics and Astronomy, University of Manitoba, Winnipeg R3T 2N2, Canada}
	\affiliation{Manitoba Quantum Institute, University of Manitoba, Winnipeg R3T 2N2, Canada}
	%
	%
	%\date{\today}
	%
	%%%%%%%%%%%%%%%%%%%%%%%%%%%%%%%%%%%%%%%%%%%%%
	%%%%%%%%%%%%%%%%%%%%%%%%%%%%%%%%%%%%%%%%%%%%%%%%%%%%%%%%%%%%%%%%%%%%%%%%%%%
	%%%%%%%%%%%%%%%%%%%%%%%%%%%%%%%%%%%%%%%%%%%%%%%%%%%%%%%%%%%%%%%%%%%%%%%%%%%
	\begin{abstract}
		Bilayer Moir\'e structures are a highly tunable laboratory to investigate the physics of strongly correlated electron systems. Moir\'e transition metal dichalcogenides at low-energies, in particular, are believed to be described by a single narrow band Hubbard model on a triangular lattice with spin-orbit coupling. Motivated by recent experimental evidence for superconductivity in twisted bilayer materials, we investigate the possible superconducting pairings in a two-dimensional single band Rashba-Hubbard model. Using a random-phase approximation in the presence of nearest and next-nearest neighbor hopping, we analyze the structure of spin fluctuations and the symmetry of the superconducting gap function. We show that Rashba spin-orbit coupling favors ferromagnetic fluctuations which strengthen triplet superconductivity. If parity is violated due to the absence of spatial inversion symmetry, singlet ($d$-wave) and triplet ($p$-wave) channels of superconductivity will be mixed. Moreover, we show that time-reversal symmetry can be spontaneously broken leading to a chiral superconducting state. Finally, we consider quasiparticle interference as a possible experimental technique to observe the superconducting gap symmetry.
		% 	%
		% 	Heterostructures are capable of realizing the coexistence of strongly correlated electron systems and broken inversion symmetry.
		% 	%
		% 	These can be considered as a platform to comprehend the coincidence of Rashba spin-orbit coupling and Hubbard interaction. 
		% 	%
		% 	Motivated by the current emergence of superconductivity in twisted bilayer materials with an effective triangular lattice, we investigate the superconducting pairing of a Rashba-Hubbard model in a two-dimensional triangular lattice.
		% 	% 
		% 	Using random-phase approximation in the presence of the first- and second-neighbor hopping, we analyze the structure of spin fluctuations and symmetry of supoerconducting gap function.
		% 	%
		% 	We show that Rashba spin-orbit coupling favors in-plane ferromagnetic fluctuations, which strengthens formation of triplet superconductivity.
		% 	% 
		% 	Since the violation of parity in the absence of spatial inversion symmetry, the singlet ($d$-wave) and triplet ($p$-wave) channels of superconductivity will be mixed together.
		% 	%
		% 	Moreover, spontaneous breaking of time-reversal symmetry in triangular lattice generates a degenerate superconducting ground state that results in emergence of chiral superconductivity.
		% 	%
		% 	Finally, we propose a possible experimental technique for observation of
		% the superconducting gap symmetry based on the quasiparticle interference approach.
		% 	%‬‬
	\end{abstract}
	%%%%%%%%%%%%%%%%%%%%%%%%%%%%%%%%%%%%%%%%%%%%%%%%%%%%%%%%%%%%%%%%%%%%%%%%%%%
	%
	\maketitle
	%
	%%%%%%%%%%%%%%%%%%%%%%%%%%%%%%%%%%%%%%%%%%%%%%%%%%%%%%%%%%%%%%%%%%%%%%%%%%%
	\section{Introduction}
	There has been a growing interest in exotic quantum phasesof twisted van der Waals bilayer materials in recent years ~\cite{Arora_2020,Kennes_Nat_Phys_2021,Das_2021,Lin_PRResearch_2021,Stern_PRB_2021,Andrei_Nature_2021,Chen_Nature_2021,Cao_Nature_2020,Huang_Nature_2017}.
	In such systems, the band structure is highly tunable by twisting one layer with respect to the other~\cite{Wu_PRL_2019,Yuan_PRB_2020,Rickhaus_SCience_2021}.
	In particular, the bandwidth can be tuned to be smaller than the magnitude of the electron interactions and, as a result, a strongly correlated electron system is formed~\cite{Cao_Sc_Nature_2018,Bernevig_PRL_2019,Vishvanath_PRL_2019,Adam_PRResearch_2020,Nori_PRB_2020}.
	By electron or hole-doping the Mott insulating state, unconventional superconducting phases can potentially be realized~\cite{Wu_PRL_2018,Cao_Nature_2018,Lu_Nature_2019,Cao_Science_2021}.
	There have been many recent theoretical and experimental reports on doped twisted bilayer graphene (TBG) that are consistent with the formation of a chiral $d+{\mi}d$-wave topological superconductivity (TSC) near half-filling~\cite{Liu_PRL_2018,Kennes_PRB_2018,Vishvanath_Nature_2019,Yankowitz_Science_2019}.
	A related Moir\'e system, which has also been studied both theoretically and experimentally in recent years, are bilayers of group-VI semiconducting transition metal dichalcogenides (TMDs) such as WSe$^{}_2$~\cite{Wu_MAc_PRL_2018,Wang_NAture_2020,Wang_NAture_2020},
	which realize an effective triangular lattice~\cite{Wu_PRL_2018,Wu_PRL_2019,Wang_NAture_2020,Regan_2020,Tang_2020,Scherer_2021}. These materials can host exotic quantum phases such as correlated insulating states, unconventional superconductors, fractional quantum Hall states, and quantum spin liquids~\cite{Wu_PRL_2018,Wu_PRL_2019,Ruiz_2019,Schrade_2019,Wang_NAture_2020,Zhou_2021,Jung_PRB_2014,Zare_2021,Scherer_2021,Kiese_arxiv_2021}.
	While graphene behaves like a Dirac semimetal with SU(2) spin rotational symmetry, TMDs are semiconductors with a large band gap and a large spin-orbit coupling (SOC)~\cite{Wu_MAc_PRL_2018,Wu_PRL_2019,Das_Sarma_PRResearch_2020}.
	Under the assumption that disorder can be ignored, TMDs therefore appear to be easier subjects for theoretical studies than TBG because of the smaller number of low-energy degrees of freedom and the presence of narrow bands for a wide range of twisting angles~\cite{Wu_MAc_PRL_2018,Wu_PRL_2019,Das_Sarma_PRResearch_2020}.
	In particular, it appears to be possible to understand many of their physical properties based on an appropriate single-band Hubbard model.
	One of the main open questions in condensed matter physics is the pairing mechanism and the symmetry of the superconducting phase in unconventional high-temperature superconductors such as the cuprates or the pnictides~\cite{Muller_Bednor_1986,Kamahira_ACS_2008,Biderang_PRB_2017,Romer_PRB_2021}. 
	There appears to be a consensus that the electron-phonon coupling is not strong enough to overcome the Coulomb interaction between the electrons in order to form Cooper pairs.
	A common denominator of these systems is that superconducting phases appear in the vicinity of magnetic instabilities and magnetic ordered phases.
	A possible candidate for a pairing mechanism in unconventional superconductors is therefore the exchange of magnons or paramagnons to form Cooper pairs~\cite{SCalapino_PRB_1986}.
	Another interesting aspect of non-BCS type superconductors is that they are capable of hosting TSC such as chiral $p$-wave and chiral $d$-wave, or helical/chiral parity-mixed states~\cite{Sato_PRB_2009,Schmalian_NAt_Com_2014,Sato_IOP_2017}. 
	%
	%{\MB
	%Coexistence of the bulk gapped states together with the protected edge modes plays a crucial role for the stability of TSC.
	%
	%This is why there are too many interests in realizing the chiral triplet superconductivity such as $p\pm {\rm i}p$, and $f\pm {\rm i}f$ pairings.}
	%
	Specifically, the twisted bilayer of TMDs with an effective triangular or honeycomb lattice can potentially produce chiral superconductivity if time-reversal symmetry (TRS) becomes spontaneously broken in their superconducting ground state. 
	%
		%%
	%%%%%%%%%%%%%%%%%%%%%%%%%%%%%%%%%%%%%%%%%%%%%%%%%%%
	%%%%%%%%%%%%%%%%%%%%====== figure ======%%%%%%%%%%%%%%%%%%%
	%%%%%%%%%%%%%%%%%%%%%%%%%%%%%%%%%%%%%%%%%%%%%%%%%%%
	\begin{figure}[t]
		\begin{center}
			\hspace{-0.65cm}
			\includegraphics[width=1.04 \linewidth]{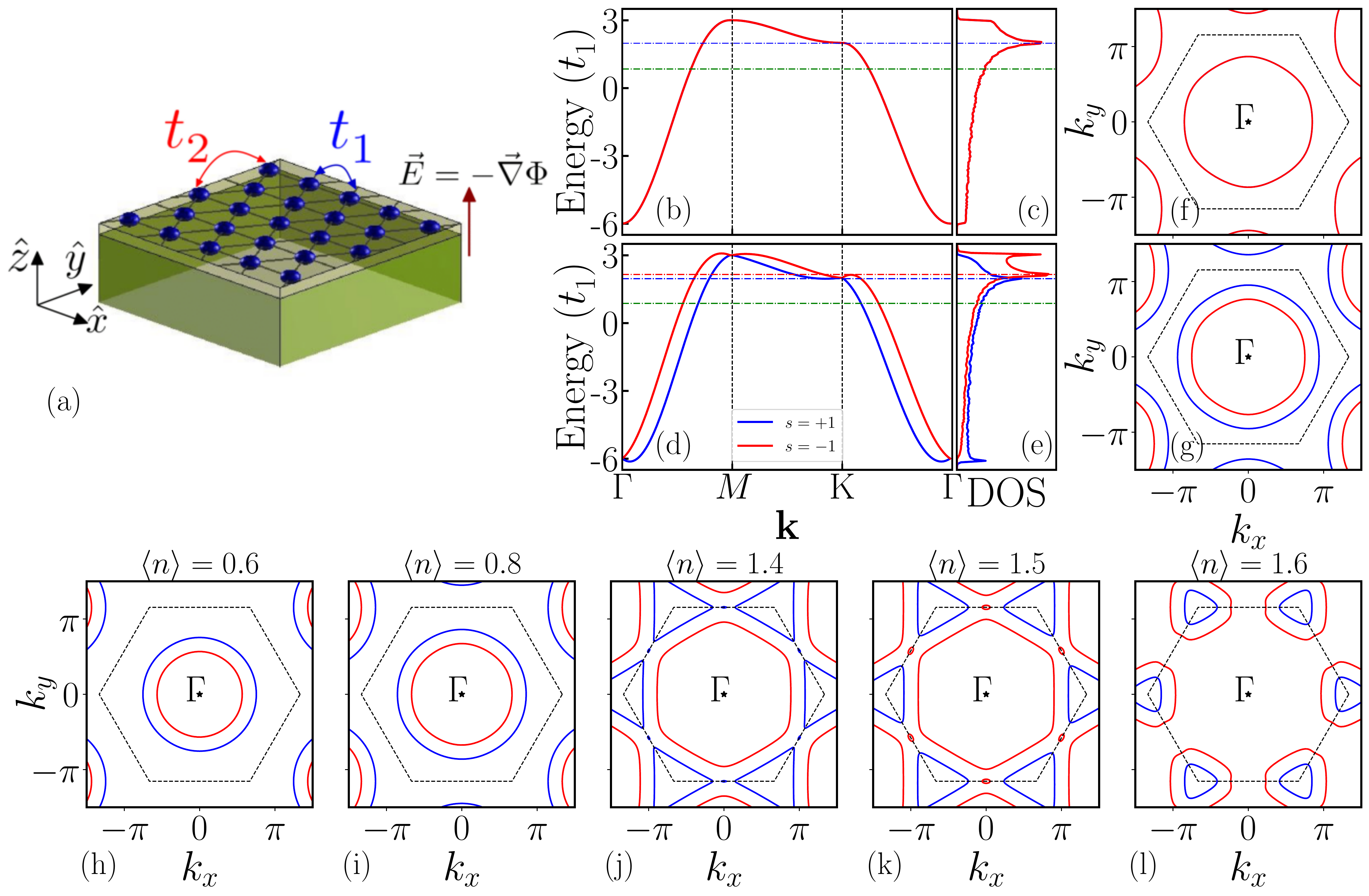} 
		\end{center}
		\caption{
			(a) Geometry of the considered system: 
			%A 2D layer of an effective triangular lattice is deposited on a semi-infinite substrate. 
			%
			The blue spheres represent ${\cal R}^{M}_{M}$ high-symmetry positions of a twisted bilayer of TMD, forming an effective triangular lattice on top of a substrate.
			$t^{}_1$ and $t^{}_2$ are the nearest- and the next-nearest- hopping, respectively.
			Broken inversion symmetry induces and effective electric field.
			%An effective external electric field is induced in the 2D layer as the result of an electric potential gradient due to broken inversion symmetry.
			%
			Also shown is the band structure of the non-interacting model for $t^{}_2=0$, for (b) $V^{}_{\rm SOC}=0$ and (d) $V^{}_{\rm SOC}=0.3$ at $T=0.03$. (c,e) The corresponding densities of state. The green line indicates half filling and the red and blue dashed lines the positions of the van Hove singularities.
			(f,g) Fermi surface for $\delta=0$ and $V_{\rm SOC}=0$ and $V_{\rm SOC}=0.3$. The central hexagons (black dashed lines) indicate the Brillouin Zone. 
			%
				%(h-l) Fermi surface as function of filling.
				%
			(h,i) In the hole-doped region, the Fermi surface consists of two hole pockets centered around the $\boldsymbol{\Gamma}$ point.
			(j,k) Close to the van Hove singularities, the Fermi surface consists of hole- and an electron-like pockets.
			(l) For large electron doping, both helical Fermi surfaces become electron pockets.
		}
		\vspace{0.73cm}
		\label{Fig:Geometry_Bands_DOS_FS}
	\end{figure}
	%%%%%%%%%%%%%%%%%%%%%%%%%%%%%%%%%%%%%%%%%%%%%%%%%%%
	%%%%%%%%%%%%%%%%%%%%%%%%%%%%%%%%%%%%%%%%%%%%%%%%%%%
	%
	
	%{\MB In addition, theoretical ~\cite{Li_2019H,David_2019,Naimer_2021} and experimental~\cite{Lin_2021} studies reveal that it is possible to realize the proximity-induced SOC such as Ising- and Rashba-SOC~\cite{Li_2019H,David_2019,Li_2019H} in Moir\'e systems.}
	%
	Parallel with the discovery of topological superconductors, progress in the fabrication of heterostructures has been made based on a wide variety of crystal growth methods, such as molecular beam epitaxy (MBE)~\cite{Chakhallin_SCience_2007,Mizukami_NAture_2011,Shimuzawa_PRL_2014,Naritrsuka_PRB_2017,Naritsuka_PRL_2018}.
	%
	%A very important and distinguishing aspect of these heterostructures is their high tunability.
	%
	A very important aspect of these structures is that inversion symmetry is, in general, broken leading to the emergence of antisymmetric Rashba or Ising SOCs~\cite{Bauer_Sigrist_NCS_Book_2012,Yanase_PRResearch_2021,Li_2019H,David_2019,Li_2019H,Li_2019H,David_2019,Naimer_2021,Lin_2021}.
	As a consequence, spin-flip scattering processes are present and can strongly affect the magnetic structure and the symmetry of the superconducting state~\cite{Tanaka_PRB_2007,Greco_PRL_2018}.
	In particular, spin is no longer a good quantum number and singlet and triplet pairings in the superconducting state will be mixed~\cite{Sigrist_PRL_2004,Fujimoto_PRB_2009,Biderang_PRB_2018}.
	The locations of the high-symmetry points in a TMD Moir\'e superlattice are denoted by ${\cal R}^{\alpha}_{\beta}$, in which $\alpha$ and $\beta$ are running over metal (M) and chalcogene (X) atoms of the top and bottom layers, respectively.
	In general, for a twisted bilayer of TMD these high-symmetry points are: ${\cal R}^M_M$, ${\cal R}^X_M$, and ${\cal R}^M_X$~\cite{Wu_PRL_2019,Das_Sarma_PRResearch_2020}.
	The structure has a D$^{}_3$ point-group symmetry for rotations around the $\hat{z}$ axis along with a twofold C$^{}_{2v}$ symmetry around the in-plane $\hat{y}$ axis~\cite{Wu_PRL_2019,Das_Sarma_PRResearch_2020}.
	The effective Moir\'e lattice of the twisted bilayer of TMDs can be easily changed by tuning the local density of states (LDOS) at high-symmetry positions using a gate voltage between the layers~\cite{Cao_Nature_2018,Cao_Sc_Nature_2018,Das_Sarma_PRResearch_2020}.
	There are two possible structures for the effective lattice of a twisted bilayer of TMDs: (i) honeycomb, or (ii) triangular lattices.
	In this paper, we will focus on the effective triangular lattice. This case is realized if the ${\cal R}^M_M$ points have a much larger LDOS than the other high-symmetry points.
	We aim to study the magnetic and superconducting instabilities in a twisted bilayer of TMD which has been deposited on top of a bulk substrate.
	This setup, sketched in Fig.~\ref{Fig:Geometry_Bands_DOS_FS}(a), will combine the physics of strongly correlated electrons with Rashba SOC.
	We note that in real twisted TMD heterostructures there is also a substrate top layer and a top gate. 
	Here we show schematically the structure without the top layer.
	In the limit of a single narrow band being relevant for the low-energy physics, we are thus trying to understand the different phases and potential pairings of a Rashba-Hubbard model on an effective triangular lattice.
	For large twist angles, which can be realized in TMDs, the Moir\'e unit cell becomes relatively small and hopping integrals between next-nearest neighbors might no longer be negligible. 
	We therefore will include such processes as well.
	We will consider both the magnetic susceptibility and a magnon-mediated pairing using the random phase approximation (RPA).
	To find the symmetry of the dominant superconducting pairing channel, we will then solve the self-consistent gap equation within BCS theory.

	Our paper is organized as follows: In Sec.~\ref{Model} we will define the considered model and calculate its non-interacting band structure and DOS for various doping levels. In Sec.~\ref{Suscept} we consider the bare and RPA spin susceptibilities as a function of hopping amplitudes, SOC, and doping level. Also within the RPA approximation, we then consider in Sec.~\ref{Pairing} the pairing of electrons, separating the effective vertex into singlet and triplet channels and neglecting its frequency dependence. In Sec.~\ref{QPI} we then discuss the response which would be seen in a quasi-particle interference experiment. We note that this scanning tunneling microscopy (STM) technique is only possible on a heterostructure without a substrate top layer and without a top gate as shown schematically in Fig.~\ref{Fig:Geometry_Bands_DOS_FS}(a) and is thus, for now, a purely theoretical consideration. In the final section we then provide a brief summary and discussion of our main results.
	
	% Here, we are going to investigate the problem of superconductivity within weak coupling approach, in which two electrons of opposite momenta on the Fermi surface (FS) attract each other through spin fluctuations and exchange of magnons.
	% %
	% We consider the magnetic structure of the systems using random phase approximation (RPA).
	% %
	% Furthermore, we solve the self-consistent superconducting gap equation within BCS theory to find the symmetry of the dominant gap function.
	% %
	% Finally, we provide the quasiparticle interference (QPI) patterns for both non-magnetic and magnetic impurity scattering to introduce an experimental method for observing the signature of superconducting gap function symmetry in the twisted bilayer of TMDs.
	%  
	
	%===========================================================
	\section{Model and non-interacting band structure}
	\label{Model}
	%===========================================================
	We start by introducing the two-dimensional Hubbard model which we will consider in the following and discuss the band structure and DOS in the non-interacting limit.
	\subsection{Rashba-Hubbard Model on the Triangular Lattice}
	Motivated by recent experiments on  twisted heterostructures which have an effective triangular Moir\'e lattice~\cite{Yankowitz_Science_2019,Cao_Nature_2018,Cao_Sc_Nature_2018,Ruiz_2019,Tang_2020,Scherer_2021,Xian_2019,Ni_2019}, we consider a two-dimensional lattice of atoms deposited on a thick substrate, see Fig.~\ref{Fig:Geometry_Bands_DOS_FS}(a).
	At small twist angles, which cause a Moir\'e pattern with an enlarged unit cell~\cite{Wu_MAc_PRL_2018}, there are hundreds of energy bands due to the high number of electrons inside the unit cell~\cite{Liu_PRL_2018}.
	However, nearly flat low-energy bands with a total bandwidth of only several meV~\cite{Liu_PRL_2018} can arise which are well isolated from the other high energy bands. To understand the low-energy physics in this situation, it thus suffices to consider only the low-energy bands.
	If these bands are partially filled, magnetic and superconducting instabilities can occur~\cite{Lu_Nature_2019,Adam_PRResearch_2020,Cao_Nature_2018}. 
	In addition, the broken inversion symmetry at the intersection of two subsystems generates a perpendicular electric field, which induces the well-known Rashba SOC~\cite{Bauer_Sigrist_NCS_Book_2012}.
	%
	%A very thin insulating layer separates the thin layer and substrate to avoid tunnelling of electrons between them.
	%
	%
	The single-orbital Rashba-Hubbard model on a two-dimensional triangular lattice is then expressed as 
	%
	%===========================================================
	%========================Equation===========================
	%===========================================================
	\begin{equation}
		{\cal H}
		=
		{\cal H}^{}_{0}
		+
		{\cal H}^{}_{\rm int},
		\label{Eq:H_total}
	\end{equation}
	%===========================================================
	%
	in which the first term denotes the non-interacting part of the Hamiltonian, including kinetic and SOC terms,
	%
	%===========================================================
	%========================Equation===========================
	%===========================================================
	\begin{equation}
		{\cal H}^{}_{0}
		=
		{\cal H}^{}_{\rm kin}
		+
		{\cal H}^{}_{\rm SOC}.
		\label{Eq:H_0}
	\end{equation}
	%===========================================================
	%
	%
	In real space, the kinetic part is given by
	%
	%===========================================================
	%========================Equation===========================
	%===========================================================
	\begin{equation}
		{\cal H}^{}_{\rm kin}
		=
		-\mu
		-t^{}_1\!\!
		\sum_{\langle ij \rangle,\sigma}
		c^{\dagger}_{i\sigma}
		c^{}_{j\sigma}
		-t^{}_2\!\!
		\sum_{\langle\!\langle ij \rangle\!\rangle,\sigma}
		c^{\dagger}_{i\sigma}
		c^{}_{j\sigma}
		+
		{\rm h.c.},
		\label{Eq:H_kin_Real}
	\end{equation}
	%===========================================================
	%
	with $\mu$, $t^{}_1$, and $t^{}_2$ being the chemical potential, and the nearest and next-nearest neighbor hopping amplitudes, respectively.
	The operator $c^{\dagger}_{i\sigma}$ ($c^{}_{i\sigma}$) creates (annihilates) an electron with spin $\sigma=\uparrow,\downarrow$ at lattice site $i$.
	Hereafter, we set $t^{}_1=1$, unless specifically mentioned otherwise. 
	As already eluded to earlier, we include a hopping $t_2$ because for large twist angles such hopping processes might no longer be negligible 
	and they can be tuned using an out-of-plane displacement field~\cite{Das_Sarma_PRResearch_2020}.
	However, $t_2$ is in general expected to be quite small.
	In momentum space, the kinetic part of the Hamiltonian
	% within the basis $\Psi^{\dagger}_{\bk}=(c^{\dagger}_{\bk\uparrow},c^{\dagger}_{\bk\downarrow})$ 
	is given by ${\cal H}^{}_{\rm kin}\!=\!\sum_{\bk,\sigma}\varepsilon^{}_{\bk}c^{\dagger}_{\bk\sigma}c^{}_{\bk\sigma}$ with dispersion 
	%
	%===========================================================
	%========================Equation===========================
	%===========================================================
	\begin{equation}
		\varepsilon^{}_{\bk}
		=
		-\mu
		-2t^{}_1
		\sum_{\gamma^{}_1=1}^{6}
		\cos(\bk \cdot \boldsymbol \br^{}_{\gamma^{}_1})
		-2t^{}_2
		\sum_{\gamma^{}_2=1}^{6}
		\cos(\bk \cdot \boldsymbol \bR^{}_{\gamma^{}_2}).
		\label{Eq:dispersion}
	\end{equation}
	%===========================================================
	%
	Here, $\br\!\!\in\!\!\lbrace 
	\pm\hat{x},
	\pm\frac{1}{2}\hat{x}\pm\frac{\sqrt{3}}{2}\hat{y} 
	\rbrace$,
	and 
	$\bR\!\!\in\!\lbrace 
	\pm\frac{3}{2}\hat{x}\pm\frac{\sqrt{3}}{2}\hat{y},
	\pm\sqrt{3}\hat{y}
	\rbrace$
	are the real space vectors connecting the nearest- and next-nearest sites, respectively.
	%
	% Moreover, subscripts $\gamma^{}_{1}$ and $\gamma^{}_{2}$ are coordination indices, running over the numbers of first and second neighbors.
	%
	Spin and orbital degrees of freedom are coupled by the SOC Hamiltonian
	%
	%===========================================================
	%========================Equation===========================
	%===========================================================
	\begin{equation}
		{\cal H}^{}_{\rm SOC}
		=
		V^{}_{\rm SOC}
		\sum_{\bk,\sigma\sigma'}
		(\bg^{}_{\bk}
		\cdot
		\hat{\boldsymbol{\sigma}})^{}_{\sigma\sigma'}
		c^{\dagger}_{\bk\sigma}
		c^{}_{\bk\sigma},
		\label{Eq:H_SOC}
	\end{equation}
	%===========================================================
	%
	in which the antisymmetric Rashba SOC (ASOC) $\bg$-vector 
	%
	%===========================================================
	%========================Equation===========================
	%===========================================================
	\begin{equation}
		\bg^{}_{\bk}=
		\Big(
		\frac{\partial\varepsilon^{}_{\bk}}{\partial k_y}
		,
		-\frac{\partial\varepsilon^{}_{\bk}}{\partial k_x}
		,0
		\Big),
		\label{Eq:SOC-g}
	\end{equation}
	%===========================================================
	% 
	is an odd function with respect to momentum and $V^{}_{\rm SOC}$ is the coupling constant.
	This type of ASOC plays the role of a momentum-dependent Zeeman field that splits the two-fold spin degenerate bands into positive and negative helical bands with energy dispersion
	%
	%===========================================================
	%========================Equation===========================
	%===========================================================
	\begin{equation}
		\xi^{}_{\bk,s}=\varepsilon^{}_{\bk}+s|\bg^{}_{\bk}|,
		\label{Eq:Energy_normal}
	\end{equation}
	%===========================================================
	%
	and helicity $s=\pm$, which is taken as a band index.
	We assume, furthermore, that the Coulomb repulsion can be truncated to an  on-site Hubbard interaction
	%
	%
	%===========================================================
	%========================Equation===========================
	%===========================================================
	\begin{equation}
		{\cal H}^{}_{\rm int}
		=
		U\sum_{i}n^{}_{i\uparrow}n^{}_{i\downarrow} \; .
		\label{Eq:H_Hubbard}
	\end{equation}
	%===========================================================
	%
	Here $U$ is the Hubbard constant and $n^{}_{i\sigma}$ the number operator at site $i$ with spin $\sigma$.
	Throughout this paper, we set $\hbar = k^{}_B=1$.

	\subsection{Non-interacting band structure}
	Let us first consider the non-interacting case.
	Fig.~\ref{Fig:Geometry_Bands_DOS_FS}(b) depicts the non-interacting band structure for the $\Gamma$MK$\Gamma$ high-symmetry path with $t^{}_2=0$ and $V^{}_{\rm SOC}=0$ while Fig.~\ref{Fig:Geometry_Bands_DOS_FS}(d) shows the band structure with $t^{}_2=0$ and $V^{}_{\rm SOC}=0.3$. 
	The bandwidth in this example is almost $W\sim 10 t^{}_1$ but can easily be tuned by changing the twist angle between the layers.
	The dashed lines indicate half filling and the positions of the van Hove singularities (VHSs), respectively. 
	Note that for $V_{\rm SOC}=0$ the bands are largely flat around the time-reversal invariant K-point leading to a type-I VHS \cite{Yao_2015}, see Fig.~\ref{Fig:Geometry_Bands_DOS_FS}(c).
	%		
	%{\MB In this case, we do not expect the stability of the spin-triplet superconductivity, since the triplet-pairing is an odd function in terms of momentum and must vanishes at type-I saddle points.}
	% 
% 	In Fig.~\ref{Fig:Geometry_Bands_DOS_FS}(e) the density of states (DOS) versus energy of the helical bands is shown, revealing the presence of VHSs for light electron-doped case.
% 	%
% 	These singularities are expected to play an important role in determining the dominant scattering channels.
	%
	In the presence of SOC, the saddle points appear along the ${\bf K-M }$ and ${\bf K-\Gamma }$ paths, away from time-reversal invariant momenta, leading to type-II VHSs as shown in Fig.~\ref{Fig:Geometry_Bands_DOS_FS}(e). These type-II VHSs are expected to lead to a competition of  singlet and triplet pairing in the superconducting phase.
	We observe that when comparing the DOS to the experimental data  in Ref.~\onlinecite{Wang_NAture_2020}, the positions of the van Hove singularities in our model seem to correspond to a non-zero displacement field in the experiment. It is not clear to us why that is the case.
	Spin-orbit coupling leads to a splitting of the Fermi surface into two helical surfaces as is shown for the half-filled case in Fig.~\ref{Fig:Geometry_Bands_DOS_FS}(f,g). For a fixed strength of SOC we can then study the evolution of the Fermi surface with doping. For fillings $\langle n \rangle=0.6,~0.8$ shown in Fig.~\ref{Fig:Geometry_Bands_DOS_FS}(h,i), both helical Fermi surfaces consist of hole-like pockets. Upon approaching the VHSs the topology of the Fermi surface changes and one of the hole pockets becomes an electron-like pocket, see Figs.~\ref{Fig:Geometry_Bands_DOS_FS}(j,k). These pockets are a direct result of the presence of proximity-induced Rashba SOC and are important for the superconducting properties of the system. Far away from the VHSs (see Fig.~\ref{Fig:Geometry_Bands_DOS_FS}(l)), both helical Fermi surface become electron-like pockets.

	\section{Spin Susceptibilities}
	\label{Suscept}
	Because we will consider spin fluctuations as the potential glue for electron pairing and superconductivity, we start by exploring the spin susceptibility as a function of next-nearest neighbor hopping, SOC, and doping.
	
	\subsection{Green's functions and RPA susceptibilities}
	The free electron Matsubara Green's function for a system in the presence of ASOC at $U=0$ is expressed by a $2\times 2$ matrix in spin space,
	%
	%===========================================================
	%========================Equation===========================
	%===========================================================
	\begin{equation}
		\hat{\cal G}^{0}_{\bk}(\mi\nu_n)
		=
		\Big[
		(
		{\mi}\nu_n 
		-
		\varepsilon^{}_{\bk}
		)
		\hat{\mathbb{I}}
		-
		V^{}_{\rm SOC}
		~{\bg}^{}_{\bk}
		\cdot
		\hat{\boldsymbol{\sigma}}
		\Big]^{-1}
		\label{Eq:Green_0}
		\vspace{0.1 cm}
	\end{equation}
	%===========================================================
	%
	where $\nu^{}_n=(2n+1)\pi T$ denote the fermionic Matsubara frequencies and $\hat{\mathbb{I}}$ is the $2\times 2$ identity matrix.
	The non-interacting Green's function in the spin basis can be mapped into the band space using the relationship
	%
	%
	%===========================================================
	%========================Equation===========================
	%===========================================================
	\begin{equation}
		\hat{\cal G}^{0}_{\bk}(\mi\nu_n)
		=
		\frac{1}{2}
		\sum_{s=\pm}
		\Big[
		\hat{\mathbb{I}}
		+s
		\hat{\bg}^{}_{\bk}
		\cdot
		\hat{\boldsymbol{\sigma}}
		\Big]
		{\cal G}^{0}_{\bk,s}(\mi\nu_n),
		\label{Eq:Green_0_spin_2_band}
	\end{equation}
	%===========================================================
	%
	in which $\hat{\bg}^{}_{\bk}={\bg}^{}_{\bk}/|{\bg}^{}_{\bk}|$, and ${\cal G}^{0}_{\bk,s}(\mi\nu_n)=
	[\mi\nu_n-\xi^{}_{\bk,s}]^{-1}_{}$ is the free electron Matsubara Green's function in band space.
	To investigate the spin fluctuations and electron instabilities of the system, we adopt the standard RPA approach.
	%
	% We begin from the normal state spin susceptibility and evaluate its renormalization because of the presence of interactions up to RPA level.
	%
	In the framework of linear response theory, the spin susceptibility at $U=0$ with momentum $\bq$ and bosonic Matsubara frequencies $\omega_m=2m\pi T$ is defined by~\cite{Cobo_PRB_2016}
	%
	%
	%===========================================================
	%========================Equation===========================
	%===========================================================
	\begin{align}
		\begin{aligned}
			\hat{\chi}^{(0)\sigma_3\sigma_4}_{\bq,\sigma_1\sigma_2}
			&({\mi}\omega_m)=
			\\
			&-\frac{T}{N}
			\sum_{\bk,{{\mi}\nu^{}_n}}
			{\cal G}^{0}_{\bk,\sigma_1\sigma_2}({\mi}\nu_n)
			{\cal G}^{0}_{\bk+\bq,\sigma_3\sigma_4}({\mi}\omega_m+{\mi}\nu_n).
			\label{Eq:Bare_Kappa}
		\end{aligned}
	\end{align}
	%===========================================================
	%
	Using Eq.~(\ref{Eq:Green_0_spin_2_band}) and performing the summation over the fermionic Matsubara frequencies ${\mi \nu^{}_n}$, the bare spin susceptibility in momentum-frequency space is given by
	%
	%===========================================================
	%========================Equation===========================
	%===========================================================
	\begin{align}
		\begin{aligned}
			\hat{\chi}^{(0)\sigma_3\sigma_4}_{\bq,\sigma_1\sigma_2}
			({\mi}\omega_m)=
			\frac{1}{4N}
			\sum_{\bk,ss'}
			\zeta^{\bk\bq,ss'}_{\sigma^{}_1\sigma^{}_2\sigma^{}_3\sigma^{}_4}
			\frac{n^{}_{f}(\xi^{}_{\bk+\bq,s'})-n^{}_{f}(\xi^{}_{\bk,s})}
			{{\mi}\omega_m+\xi^{}_{\bk,s}-\xi^{}_{\bk+\bq,s'}},
			\label{Eq:Bare_Kappa_Explicit}
		\end{aligned}
	\end{align}
	%===========================================================
	%
	which $n^{}_{f}(\xi^{}_{\bk,s})=[1+\exp(\xi^{}_{\bk,s}/T)]^{-1}$ denotes the Fermi distribution function.
	Here, the weight factor ${\zeta}^{\bk\bq,ss'}_{\sigma^{}_1\sigma^{}_2\sigma^{}_3\sigma^{}_4}$ is obtained by~\cite{Biderang_arxiv_2019}
	%
	%===========================================================
	%========================Equation===========================
	%===========================================================
	\begin{align}
		\begin{aligned}
			\zeta^{\bk\bq,ss'}_{\sigma^{}_1\sigma^{}_2\sigma^{}_3\sigma^{}_4}
			=
			\Big[
			\hat{\mathbb{I}}
			+
			s \hat{\bg}^{}_{\bk}
			\cdot
			\hat{\boldsymbol{\sigma}}
			\Big]^{}_{\sigma_1\sigma_2}
			\Big[
			\hat{\mathbb{I}}
			+
			s' \hat{\bg}^{}_{\bk+\bq}
			\cdot
			\hat{\boldsymbol{\sigma}}
			\Big]^{}_{\sigma_3\sigma_4}.
			\label{Eq:Weight_Factor}
		\end{aligned}
	\end{align}
	%===========================================================
	%
	Next, we include the effects of the on-site Hubbard interaction by perturbatively renormalizing the bare susceptibility. Within RPA, the elements of the dressed spin susceptibility matrix are given by the following Dyson equation
	%
	%===========================================================
	%========================Equation===========================
	%===========================================================
	\begin{align}
		\begin{aligned}
			\hat{\chi}^{\sigma_3\sigma_4}_{\sigma_1\sigma_2}
			=
			\hat{\chi}^{(0)\sigma_3\sigma_4}_{\sigma_1\sigma_2}
			+
			\sum_{\lbrace\alpha_i\rbrace}
			\hat{\chi}^{(0)\alpha_1\alpha_2}_{\sigma_1\sigma_2}
			\hat{U}^{\alpha_3\alpha_4}_{\alpha_1\alpha_2}
			\hat{\chi}^{\sigma_3\sigma_4}_{\alpha_3\alpha_4},
			\label{Eq:RPA_Kappa}
		\end{aligned}
	\end{align}
	%===========================================================
	%
	where $\alpha^{}_i$ represents spin indices on the internal lines of the Feynman diagrams.
	Furthermore, $\hat{U}$ is the matrix of the bare electron-electron interactions in spin space, whose non-zero elements are  $\hat{U}^{\downarrow\downarrow}_{\uparrow\uparrow}=
	\hat{U}^{\uparrow\uparrow}_{\downarrow\downarrow}=-U$, and
	$\hat{U}^{\uparrow\downarrow}_{\downarrow\uparrow}=
	\hat{U}^{\downarrow\uparrow}_{\uparrow\downarrow}=+U$.
	It is worth to emphasize here again that in the presence of ASOC spin-flip scattering is present. Thus, both the bubble (screening) and ladder (exchange) diagrams need to be summed up to obtain the RPA spin susceptibility. 
	The longitudinal and transverse components of the dressed spin susceptibility are $\chi^{\perp}_{}=\sum_{\sigma}(\hat{\chi}^{\sigma\sigma}_{\sigma\sigma}-\hat{\chi}^{\sigma\bar{\sigma}}_{\sigma\bar{\sigma}})$, and $\chi^{\parallel}_{}=\sum_{\sigma} \hat{\chi}^{\sigma\sigma}_{\bar{\sigma}\bar{\sigma}}$, respectively.
	The former corresponds to the bubble diagrams while the latter is related to the ladder diagrams.
	\subsection{Results}
\begin{figure}[t]
		\begin{center}
			\hspace{-0.65cm}
			\includegraphics[width=1.07 \linewidth]{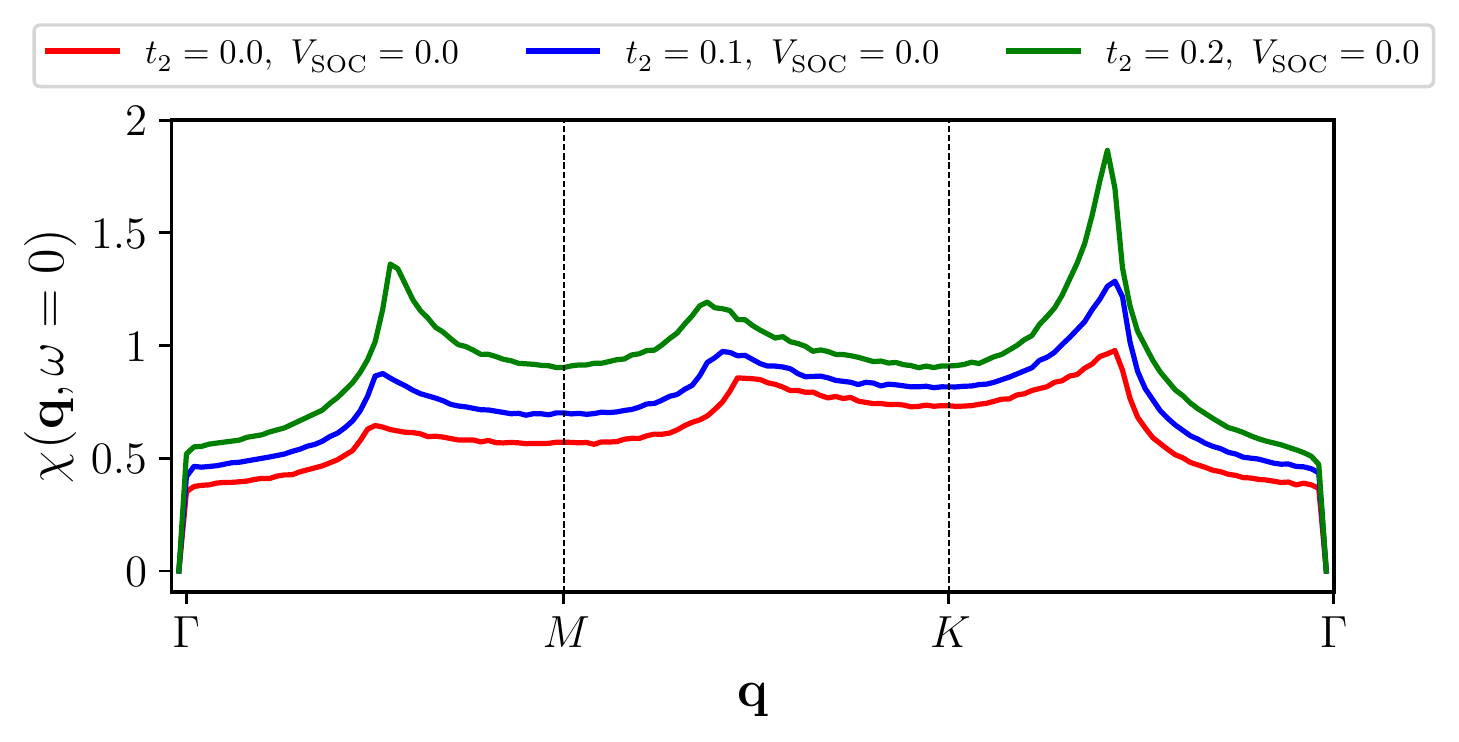} 
		\end{center}
		\vspace{-0.5 cm}
		\caption{ 
			Static ($\omega=0$) RPA spin susceptibilities in the absence of SOC at zero doping.
			%
			% 		The red, blue, and green curves correspond to $t^{}_2=0$, $0.3$, and $0.6$, respectively.
			%
			Since $V^{}_{\rm SOC}=0$, the longitudinal and transverse components of the spin susceptibility are identical.
		}
		\vspace{0.73cm}
		\label{Fig:Kappa_t2_SOC_00}
	\end{figure}

	After this brief review of the RPA formalism, we present in the following results for the static ($\omega=0$) spin susceptibility. Throughout this section, we keep the interaction strength $U=1.5$ and temperature $T=0.03$ fixed.
	In Fig.~\ref{Fig:Kappa_t2_SOC_00}, the influence of the next-nearest neighbor hopping $t_2$ on the spin structure in the absence of spin-orbit coupling is shown.
	%
	%%
	%%%%%%%%%%%%%%%%%%%%%%%%%%%%%%%%%%%%%%%%%%%%%%%%%%%
	%%%%%%%%%%%%%%%%%%%%====== figure ======%%%%%%%%%%%%%%%%%%%
	%%%%%%%%%%%%%%%%%%%%%%%%%%%%%%%%%%%%%%%%%%%%%%%%%%%
	%%%%%%%%%%%%%%%%%%%%%%%%%%%%%%%%%%%%%%%%%%%%%%%%%%%
	%%%%%%%%%%%%%%%%%%%%%%%%%%%%%%%%%%%%%%%%%%%%%%%%%%%
	%
	Note that in this case SU(2) spin-rotational symmetry is preserved. 
	For $t^{}_2=0$, the peak of the  spin susceptibility occurs in the $\boldsymbol{\Gamma}-\bK$ path near the $\bK$ point.
	Increasing the amplitude of the next-nearest neighbor hopping leads to a warping of the Fermi surface, and the position of the maximum is slightly shifted towards the $\bK$ point.
	The larger $t^{}_2$ is, the stronger the spin fluctuations in the $\boldsymbol{\Gamma}-\bK$ path are with the secondary peak shifting towards the $\bM$ point.
	In all cases, the spin fluctuations are incommensurate and for $U=1.5$ the system is not magnetically ordered.
	Moreover, for all values of the next-nearest neighbor hopping, the amplitude of the spin susceptibility at the $\boldsymbol{\Gamma}$ point is almost zero.
	I.e., pure ferromagnetic spin fluctuations are absent.
	%
	%{\HZ However, it is worthy to mention that the ferromagnetic fluctuations occur only near to VHSs due to Stoner ferromagnetic~\cite{Romer_PRB_2015,Kamogawa_2019}.
	% and quickly for filling away from VHSs. 
	%
	%The Stoner ferromagnetic criterion can be satisfied when perfect nesting of the Fermi-surface does not occur.}
	%
	% Since the maxima of spin susceptibility do not occur at the positions of high symmetry points in BZ, the magnetic fluctuations of the systems have incommensurate texture.
	% %
	% In the superconducting state, this situation corresponds to the occurrence of both spin singlet and triplet superconductivity.
	% %
	% However, because of very small ferromagnetic fluctuations, the amplitude of triplet component is very lower than the magnitude of singlet one.
	%
	
\begin{figure}[t]
		\begin{center}
			\hspace{-0.65 cm}
			\includegraphics[width=1.0 \linewidth]{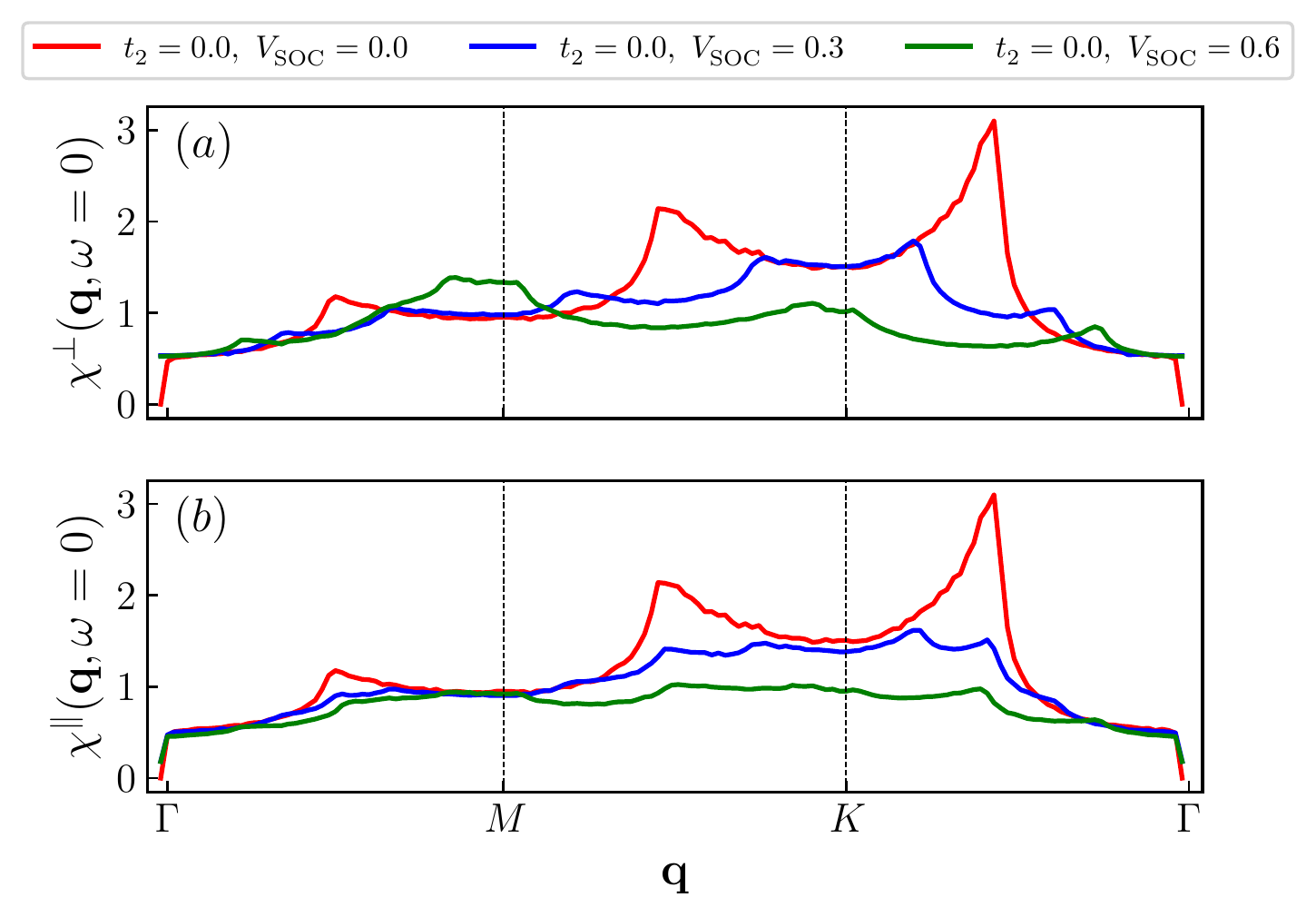} 
		\end{center}
		\vspace{-0.5 cm}
		\caption{ 
			Static ($\omega=0$) longitudinal (upper panel) and transverse (lower panel) spin susceptibilities in RPA for $t^{}_1=1$, and $t^{}_2=0$ at zero doping with $U=1.5$.
			The red, blue, and green lines represent $V^{}_{\rm SOC}=0$, $0.3$, and $0.6$, respectively.
			Note that for $V^{}_{\rm SOC}=0$, the longitudinal and transverse components of the spin susceptibility are the same due to SU$(2)$ symmetry.
		}
		\vspace{0.73cm}
		\label{Fig:Kappa_SOC_t2_00}
	\end{figure}

	Next, we show in Fig.~\ref{Fig:Kappa_SOC_t2_00} how the spin orbit coupling $V_{\rm SOC}$ influences the spin fluctuations in the absence of next-nearest neighbor hopping, $t_2=0$, and at zero doping, $\delta=0$.
	%%
	%%%%%%%%%%%%%%%%%%%%%%%%%%%%%%%%%%%%%%%%%%%%%%%%%%%
	%%%%%%%%%%%%%%%%%%%%====== figure ======%%%%%%%%%%%%%%%%%%%
	%%%%%%%%%%%%%%%%%%%%%%%%%%%%%%%%%%%%%%%%%%%%%%%%%%%
	%%%%%%%%%%%%%%%%%%%%%%%%%%%%%%%%%%%%%%%%%%%%%%%%%%%
	%%%%%%%%%%%%%%%%%%%%%%%%%%%%%%%%%%%%%%%%%%%%%%%%%%%
	%
	%
	For $V_{\rm SOC}\neq 0$, the SU(2) spin-rotational symmetry is broken and transverse and longitudinal components are different. With increasing spin-orbit coupling, we find that the spin susceptibility is, overall, suppressed. This reduction can be understood as follows:
	Rashba SOC leads to spin-flip processes and breaks the magnetic susceptibility into parallel and opposite spin channels destabilizing magnetic order. The exception are ferromagnetic fluctuations which are not affected by spin-flip scattering so that $V_{\rm SOC}$ can even increase the magnitude of the susceptibility at the $\Gamma$ point. For spin-mediated superconducting pairing we might therefore expect that triplet pairing is enhanced~\cite{Aoki_PRB_2004}.
	This point is consistent with the results of Ref.~\cite{Tanaka_PRB_2007}.

\begin{figure}[t]
		\begin{center}
			\hspace{-0.7cm}
			\includegraphics[width=1.07 \linewidth]{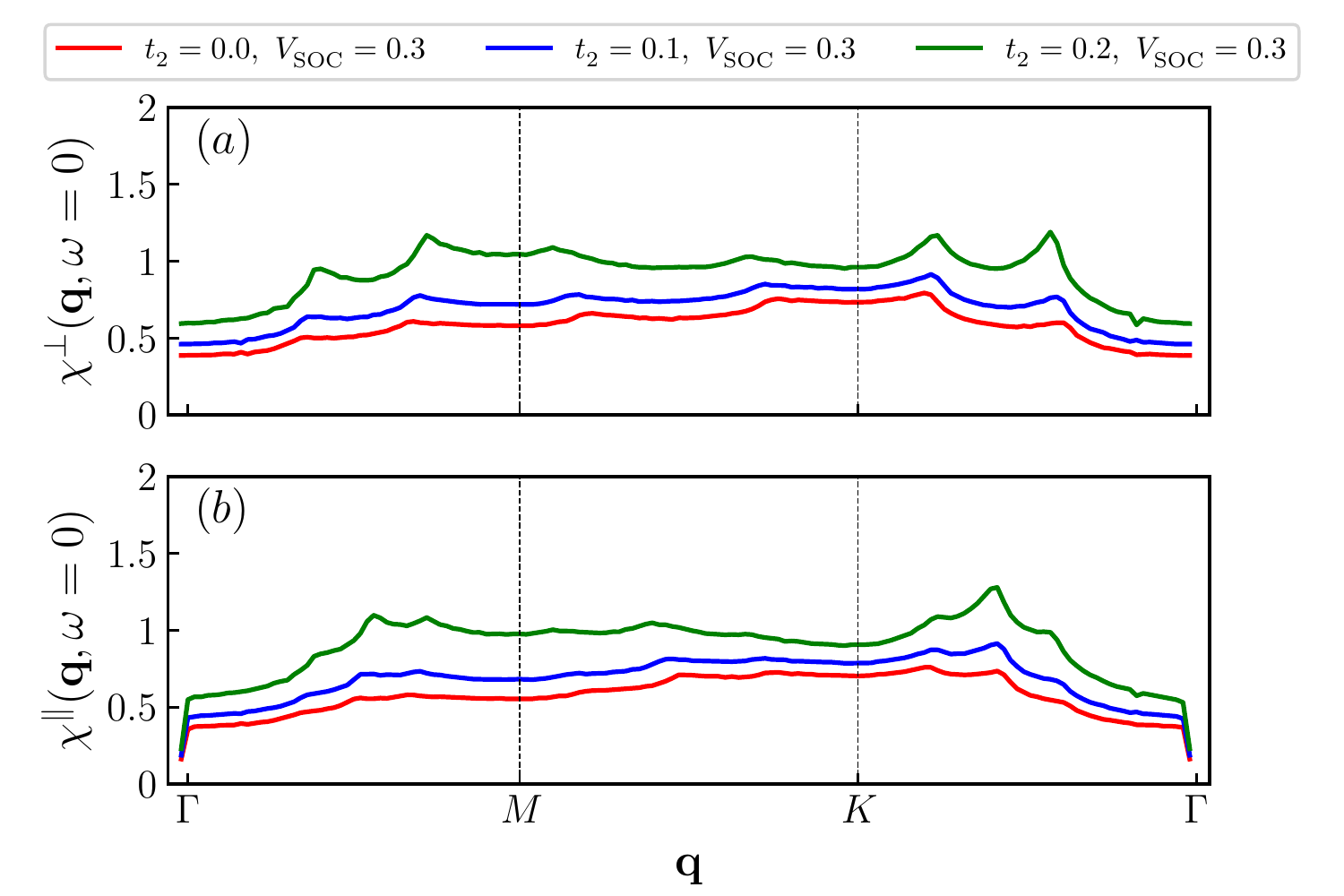} 
		\end{center}
		\vspace{-0.5 cm}
		\caption{ 
			Static ($\omega=0$) longitudinal and transverse spin susceptibilities in RPA for $V^{}_{\rm SOC}=0.3$ at zero doping for different $t_2$.
			%
			% 		The first to third rows represent the cases with $t^{}_{2}=0$, $0.3$, and $0.6$, respectively.
			% 
		}
		\vspace{0.73cm}
		\label{Fig:Kappa_t2_SOC_03}
	\end{figure}

	If the magnitude of $t_2$ is increased for a fixed strength of SOC, then magnetic fluctuations are enhanced which ultimately will bring the system for large $t_2$ close to a spin density wave (SDW) state, see Fig.~\ref{Fig:Kappa_t2_SOC_03}.
		%%
	%%%%%%%%%%%%%%%%%%%%%%%%%%%%%%%%%%%%%%%%%%%%%%%%%%%
	%%%%%%%%%%%%%%%%%%%%====== figure ======%%%%%%%%%%%%%%%%%%%
	%%%%%%%%%%%%%%%%%%%%%%%%%%%%%%%%%%%%%%%%%%%%%%%%%%%
	%%%%%%%%%%%%%%%%%%%%%%%%%%%%%%%%%%%%%%%%%%%%%%%%%%%
	%
	%
	Ferromagnetic fluctuations are also enhanced which can potentially favour a spin mediated pairing in the triplet channel. Note, however, that all the fluctuations remain incommensurate with the lattice. 
	%
	% Fig.~\ref{Fig:Kappa_t2_SOC_03} carries the information about coexistence of the second-neighbor hopping and SOC, including the effect of change in the value of $t^{}_2$.
	% %
	% In the presence of Rashba SOC, an increase in the second-neighbor hopping integral gives rise to amplifying the spin fluctuations.
	% %
	% This will ultimately brings the system into the vicinity of SDW state.
	% %
	% All the magnetic fluctuations have incommensurate behavior, deriving an admixture of singlet and triplet superconductivity.
	% %
	% However, as one can see, in the out-of-plane component of susceptibility, rising Rashba SOC increases the fluctuations at $\boldsymbol{\Gamma}$, which results in ferromagnetic fluctuations and enhancement of triplet superconductivity.
	%
	%
	
\begin{figure}[t]
		\begin{center}
			\hspace{-0.7cm}
			\includegraphics[width=1.07 \linewidth]{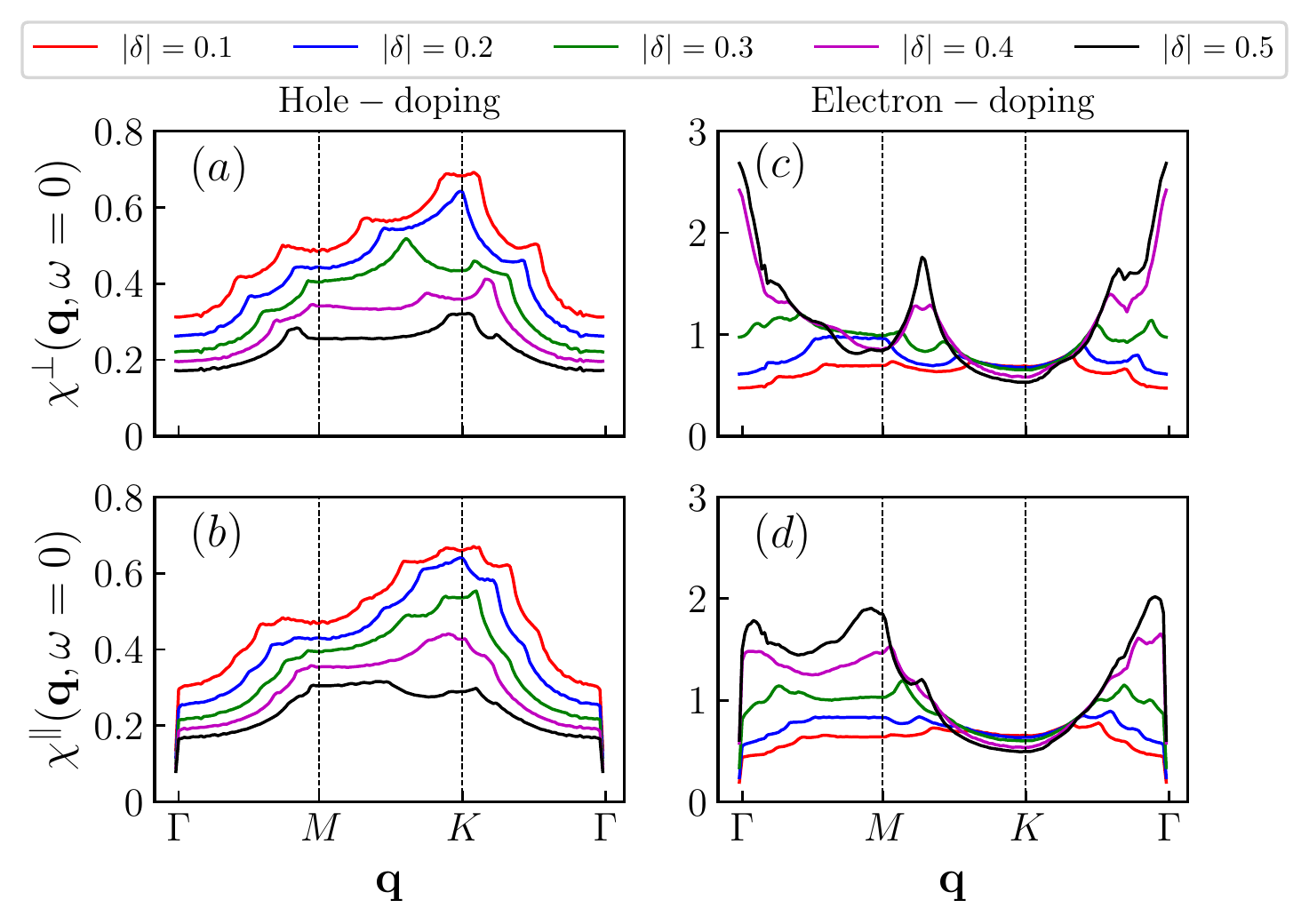} 
		\end{center}
		\vspace{-0.5 cm}
		\caption{ 
			Static ($\omega=0$) longitudinal and transverse spin susceptibilities in RPA for $t^{}_2=0$ and $V^{}_{\rm SOC}=0.3$ at different levels of hole (left column) and electron (right column) doping.
		}
		\vspace{0.2 cm}
		\label{Fig:Kappa_doping}
	\end{figure}

	%
	Finally, we also consider the effect of hole versus electron doping, see Fig.~\ref{Fig:Kappa_doping}.
	%%
	%%%%%%%%%%%%%%%%%%%%%%%%%%%%%%%%%%%%%%%%%%%%%%%%%%%
	%%%%%%%%%%%%%%%%%%%%====== figure ======%%%%%%%%%%%%%%%%%%%
	%%%%%%%%%%%%%%%%%%%%%%%%%%%%%%%%%%%%%%%%%%%%%%%%%%%
	%%%%%%%%%%%%%%%%%%%%%%%%%%%%%%%%%%%%%%%%%%%%%%%%%%%
	%%%%%%%%%%%%%%%%%%%%%%%%%%%%%%%%%%%%%%%%%%%%%%%%%%%
	%
	In the hole doped case, an increase of the doping level simply leads to an overall suppression of spin fluctuations both in the longitudinal and in the transverse component. 
	The evolution of the susceptibilities is more intricate in the electron-doped case because of the presence of a van-Hove singularity, see Fig.~\ref{Fig:Geometry_Bands_DOS_FS}. 
	For doping levels close to the van-Hove singularity the topology of the Fermi surface changes and a sharp ferromagnetic peak develops at the $\boldsymbol{\Gamma}$ point.
	This result is consistent with Ref.~\cite{Greco_PRB_2020}.
	If the strength of the on-site Hubbard interaction is increased further, the system will therefore be driven into a ferromagnetically ordered state---at least at the RPA level of approximation.
	Close to this ferromagnetic state we might expect the potential for an enhanced triplet pairing. Away from the van-Hove singularities the spin fluctuations are always incommensurate and short ranged.
	% the effect of charge carrier concentration on the magnetic texture of the system is shown.
	% %
	% Here, we set $t^{}_2=V^{}_{\rm SOC}=0.3$.
	% %
	% The left and right columns represent the hole- and electron-doped cases, respectively.
	% %
	% Based on the spin susceptibility data, it can be easily seen that the systems shows large magnetic fluctuations, especially at specific values of fillings.
	% %
	% It can be concluded that the properties of magnetic fluctuations are strongly filling dependent.
	% %
	% For these specific values of $t^{}_2$ and $V^{}_{\rm SOC}$, a van Hove singularity occurs between $\delta=0.2$, and $0.3$.
	% %
	% It is observed that near the van Hove singularity, a very sharp Ferromagnetic peak occurs in the spin susceptibility at $\boldsymbol{\Gamma}$ point that strongly depends on the topology of FS and the structure of magnetic susceptibility.
	% %
	% At this level of doping, increasing the Hubbard interaction $U$ enhances the amplitude of magnetic fluctuations into driving a spin density wave (SDW) order in the system. 
	% %
	% For other levels of doping, system is still far from magnetically ordered phase and show short-range magnetic fluctuations.
	% % 
	% Moreover, for most levels of doping, the texture of spin fluctuations is incommensurate, which results in an admixture of singlet and triplet superconductivity. 
	
		%
	
	\section{Superconductivity}
	\label{Pairing}
	Having obtained the spin susceptibilities we are now in a position to consider the pairing of electrons and the symmetries of the superconducting gap function.
	
	\subsection{Pairing within RPA spin fluctuation theory}
	In the channel of Cooper pairs, we consider a pair of electrons with momenta and spins $(\bk'\sigma^{}_3,-\bk'\sigma^{}_4)$ which interact and scatter with new momenta and spins $(\bk\sigma^{}_1,-\bk\sigma^{}_2)$.
	Within RPA, the interaction between the electrons involved in the scattering process will be described by
	%
	%
	%===========================================================
	%========================Equation===========================
	%===========================================================
	\begin{equation}
		{\cal H}^{\rm RPA}_{\rm int}(\bk,\bk')=
		\frac{1}{N}\!\!
		\sum_{\bk\bk',\lbrace \sigma_i \rbrace}
		\hat{\Gamma}^{\sigma_3\sigma_4}_{\sigma_1\sigma_2}(\bk,\bk')
		c^{\dagger}_{\bk\sigma_1}
		c^{\dagger}_{-\bk\sigma_2}
		c^{}_{-\bk'\sigma_4}
		c^{}_{\bk'\sigma_3}.
		\label{Eq:Effective_Int}
	\end{equation}
	%===========================================================
	%
	%
	In the presence of ASOC, the effective vertex $\hat{\Gamma}^{\sigma_3\sigma_4}_{\sigma_1\sigma_2}(\bk,\bk')$ includes the contributions of longitudinal (screening) and transverse (exchange) interactions.
	In order to find the symmetries of the superconducting instabilities, one can project the superconducting gap function into separate singlet and triplet channels. 
	In the singlet channel, the vertex function is symmetric with respect to $\bk$ and $\bk'$, and is defined by 
	%
	%===========================================================
	%========================Equation===========================
	%===========================================================
	\begin{equation}
		\hat{\Gamma}^{\rm S}_{\sigma_1\sigma_2\sigma_3\sigma_4}(\bk,\bk')=
		\frac{1}{2}
		\Big[
		\hat{U}
		+
		\frac{3}{2} \hat{U} \hat{\chi}^{}_{\bk-\bk'} \hat{U}
		+
		\frac{3}{2} \hat{U}  \hat{\chi}^{}_{\bk+\bk'} \hat{U}
		\Big],
		\label{Eq:Vertex_Function_Singlet}
	\end{equation}
	%===========================================================
	%
	while, in the triplet channel it is expressed as
	%
	%===========================================================
	%========================Equation===========================
	%===========================================================
	\begin{equation}
		\hat{\Gamma}^{\rm T}_{\sigma_1\sigma_2\sigma_3\sigma_4}(\bk,\bk')=
		-\frac{1}{2}
		\Big[
		\hat{U} \hat{\chi}^{}_{\bk-\bk'} \hat{U}
		-
		\hat{U}  \hat{\chi}^{}_{\bk+\bk'} \hat{U}
		\Big],
		\label{Eq:Vertex_Function_triplet}
	\end{equation}
	%===========================================================
	%
	which has been antisymmetrized.
	It should be mentioned that we only consider the static form of the vertex function ($\omega=0$); any frequency-dependence is neglected.
	Considering only the intra-band Cooper pairings within band basis, one can use the following transformation to find the effective pairing interaction between two electrons on the Fermi surface
	%
		%
	%===========================================================
	%========================Equation===========================
	%===========================================================
	\begin{eqnarray}
	\label{Eq:Eff_Int_Band}
	&& V^{{\rm S}/{\rm T}}_{ss'}(\bk,\bk') \\
	&&=
	\sum_{\lbrace \sigma_i \rbrace}
	\hat{\Gamma}^{{\rm S}/{\rm T}}_{\sigma_1\sigma_2\sigma_3\sigma_4}(\bk,\bk')
	\Lambda^{s*}_{\bk\sigma_1}
	\Lambda^{s*}_{-\bk\sigma_2}
	\Lambda^{s'}_{-\bk'\sigma_4}
	\Lambda^{s'}_{\bk'\sigma_3}, \nonumber
	\end{eqnarray}
	%===========================================================
	%
	with $\Lambda^{s}_{\bk\sigma}=\bra\bk,\sigma|\bk,s\ket$ connecting the state $|\bk,\sigma \ket$ in spin space with $|\bk,s \ket$ in the band basis.
	%%%%%%%%%%%%%%%%%%%%%%%%%%%%%%%%%%%%%%%%%%%%%%%%%%%
	%
	
	%
	Using Eq.~(\ref{Eq:Eff_Int_Band}) as the effective interaction within the BCS theory of superconductivity, the superconducting gap function in the singlet and triplet channels for band $s$ is then given by~\cite{Mahan_ManyBody_2000}
	%
	%
	%===========================================================
	%========================Equation===========================
	%===========================================================
	\begin{equation}
		\Delta^{}_{\bk s}
		=
		-\frac{1}{N}
		\sum_{\bk' s'}
		V^{{\rm S}/{\rm T}}_{ss'}
		\frac{\Delta^{}_{\bk' s'}}{2E^{}_{\bk' s'}}
		\tanh
		\Big(
		\frac{E^{}_{\bk' s'}}{2k_B T}
		\Big),
		\label{Eq:SC_gap_Self_Consistence}
	\end{equation}
	%===========================================================
	%
	where $E^{}_{\bk s}=\sqrt{\xi^{2}_{\bk s}+\Delta^{2}_{\bk s}}$
	%
	%
	%===========================================================
	%========================Equation===========================
	%===========================================================
% 	\begin{equation}
% 		%
% 		E^{}_{\bk s}=\sqrt{\xi^{2}_{\bk s}+\Delta^{2}_{\bk s}}
% 		%
% 		\label{Eq:SC_dispersion}
% 	\end{equation}
	%===========================================================
	%
	defines the energy dispersion of the superconducting quasiparticles.
	Close to the superconducting critical temperature ($T\rightarrow T_c$), the quasiparticle gap can be neglected leading to $E^{}_{\bk s}=|\xi^{}_{\bk s}|$. 
	The linearized gap equation can then be reduced to an eigenvalue problem to determine the leading and sub-leading superconducting instabilities of the system~\cite{Liu_PRL_2018,Romer_PRB_2021}
	%
	%===========================================================
	%========================Equation===========================
	%===========================================================
	\begin{equation}
		\lambda\Delta^{}_{\bk s}
		=
		-\frac{1}{(2\pi)^2}
		\sum_{s'}
		\oint^{}_{\rm FS}
		\frac{dk'_{\parallel}}{v^{\rm F}_{\bk' s'}}
		V^{{\rm S}/{\rm T}}_{ss'}
		\Delta^{}_{\bk' s'}.
		\label{Eq:SC_gap_Eigenvalue}
	\end{equation}
	%===========================================================
	%
 
	%
	In this equation, $dk'^{}_{\parallel}$ is the tangential component of the momentum differential on the Fermi surface and $v^{\rm F}_{\bk s}=|\nabla \xi^{}_{\bk s}|$ is the Fermi velocity of band $s$.
	The largest eigenvalue of Eq.~(\ref{Eq:SC_gap_Eigenvalue}) specifies the leading superconducting order parameter with $T_c\propto \exp(-1/\lambda)$.
	In addition, the normalized eigenvector corresponding to the largest eigenvalue determines the momentum dependence of the superconducting gap function and its nodal structure.
	Besides, we can decompose the gap function into an amplitude $\Delta^{}_0$ and a dimensionless symmetry gap function $\phi^{}_{\bk}$ for all possible superconducting pairings as $\Delta^{}_{\bk}=\Delta^{}_0 \phi^{}_{\bk}$.
	The triangular lattice has $D^{}_6$ point group symmetry, containing four one dimensional (1D) ($A^{}_1$, $A^{}_2$, $B^{}_1$, $B^{}_2$), and two 2D ($E^{}_1$, $E^{}_2$) irreducible representations (Irreps.).
	$A^{}_1$, $A^{}_2$, and $E^{}_2$ generate Cooper pairs in the singlet channel, and $B^{}_1$, $B^{}_2$, $E^{}_1$ give rise to pairing in the triplet channel~\cite{Rachel_PRB_2018}.
	The corresponding momentum dependency of gap functions ($\phi^{}_{\bk}$) for each irrep. are shown in table.~\ref{Tab:SC_gap}.
	%
			%
	%%%%%%%%%%%%%%%%%%%%%%%%%%%%%%%%%%%%%%%%%%%%%%%%%%%
	%%%%%%%%%%%%%%%%%%%%====== figure ======%%%%%%%%%%%%%%%%%%%
	%%%%%%%%%%%%%%%%%%%%%%%%%%%%%%%%%%%%%%%%%%%%%%%%%%%
	\begin{figure}[t]
		\begin{center}
			\hspace{-0.65cm}
			\includegraphics[width=1.04 \linewidth]{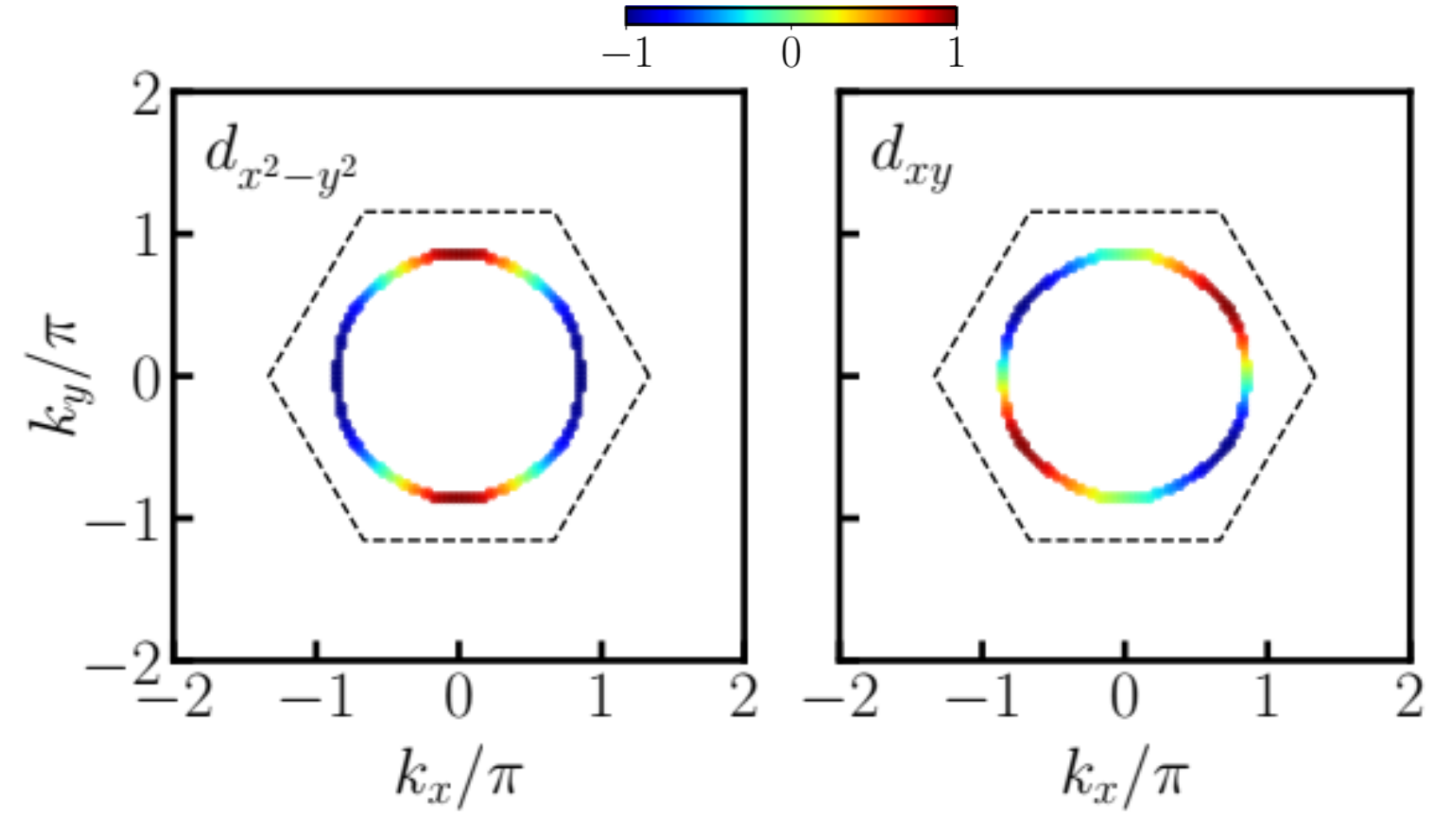} 
		\end{center}
		\vspace{-0.5 cm}
		\caption{ 
			The normalized superconducting gap functions with (left) $d^{}_{x^2-y^2}$, and (right) $d^{}_{xy}$-wave symmetries for $t^{}_1=1$, $t^{}_2=0$, and $V^{}_{\rm SOC}=0$, at doping level $\delta=0$.
			%
			% 		The absence of Rashba SOC leads to a superconducting instability in the singlet channel.
			%
			The system is expected to pick one of the degenerate superconducting ground states $d^{}_{x^2-y^2}\pm {\mi}d^{}_{xy}$ leading to a spontaneously broken time reversal symmetry.
		}
		\vspace{0.73cm}
		\label{Fig:SC_gap_FS_t2_03_SOC_0}
	\end{figure}
	%%%%%%%%%%%%%%%%%%%%%%%%%%%%%%%%%%%%%%%%%%%%%%%%%%%
	%%%%%%%%%%%%%%%%%%%%%%%%%%%%%%%%%%%%%%%%%%%%%%%%%%%
	%
	%===========================================================
	%========================Equation===========================
	%===========================================================
	\begin{table}[ht]
		\centering
		\caption{Character table of the superconducting gap functions for different irreducible representations of the point group $D^{}_6$~\cite{Rachel_PRB_2018}.}
		\vspace{0.3 cm}
		\scalebox{1.1}{
			\begin{tabular}{c c c c} 
				\hline
				Irrep. & Symmetry & $\phi^{}_{\bk}$ \\ [1ex] 
				\hline
				$A_1$ & ext.$s$-wave & $\cos k_x + 2 \cos \frac{k_x}{2} \cos \frac{\sqrt{3} k_y}{2}$
				\\ 
				$B_1$ & $f^{}_{x(x^2-3y^2)}$-wave & $\sin k_x - 2\sin \frac{k_x}{2} \cos \frac{\sqrt{3} k_y}{2}$
				\\
				$E_1$ & $p^{}_x$-wave & $\sin k_x + 2\sin \frac{k_x}{2} \cos \frac{\sqrt{3} k_y}{2}$ 
				\\
				$E_1$ & $p^{}_{y}$-wave & $ \cos \frac{k_x}{2} \sin \frac{\sqrt{3} k_y}{2}$
				\\
				%				%
				$E_2$ & $d^{}_{x^2-y^2}$ & $\cos k_x - 2 \cos \frac{k_x}{2} \cos \frac{\sqrt{3} k_y}{2}$
				\\
				$E_2$ & $d^{}_{xy}$ & $ \sin \frac{k_x}{2} \sin \frac{\sqrt{3} k_y}{2}$ 
				\\	 
				\hline
		\end{tabular}}
		\label{Tab:SC_gap}
	\end{table}
	%===========================================================
	%
	For each irrep., the eigenvalue $\lambda$ of Eq.~(\ref{Eq:SC_gap_Eigenvalue}) forms a $2\times 2$ matrix ($\lambda^{ss'}_{l}$).
	The largest eigenvalue of $\lambda^{ss'}_{l}$ gives~\cite{Romer_PRB_2015}
	%
	%
	%===========================================================
	%========================Equation===========================
	%===========================================================
	\begin{equation}
		\lambda^{ss'}_{l}
		=
		-\frac
		{
			\int^{}_{{\rm FS}^{}_s}
			\frac{dk^{}_{\parallel}}{v^{\rm F}_{\bk s}}
			\int^{}_{{\rm FS}^{}_{s'}}
			\frac{dk'^{}_{\parallel}}{v^{\rm F}_{\bk' s'}}
			\phi^{l}_{\bk}
			V^{{\rm S}/{\rm T}}_{ss'}(\bk,\bk')
			\phi^{l}_{\bk'}
		}
		{2\pi^2
			\int^{}_{{\rm FS}^{}_{s'}}
			\frac{dk'^{}_{\parallel}}{v^{\rm F}_{\bk' s'}}
			[\phi^{l}_{\bk'}]^2
		},
		\label{Eq:SC_Lambda_ss'}
	\end{equation}
	%===========================================================
	% 
	where $l$ is running over the allowed irreps. for point group $D^{}_{6}$.
	The diagonal and off-diagonal elements of matrix $\lambda^{ss'}_{l}$ relate to the Cooper pairing between to electrons in the same and opposite helical bands, respectively.
	The dominant superconducting order parameter in singlet and triplet channels for irrep. $l$ are given by the largest eigenvalue of the $\lambda^{ss'}_{l}$ matrix, which is denoted as $\lambda^{\rm eff}_{l}$.
	It is possible to estimate the superconducting critical temperature for each irrep. by $T^{l}_c=1.13\omega^{}_c \exp(-1/\lambda^{\rm eff}_{l})$, in which $\omega^{}_c$ denotes the cut-off energy and is of the order of the bandwidth. 
	%

	%%
	%%%%%%%%%%%%%%%%%%%%%%%%%%%%%%%%%%%%%%%%%%%%%%%%%%%
	%%%%%%%%%%%%%%%%%%%%====== figure ======%%%%%%%%%%%%%%%%%%%
	%%%%%%%%%%%%%%%%%%%%%%%%%%%%%%%%%%%%%%%%%%%%%%%%%%%
	\begin{figure}[t]
		\begin{center}
			\hspace{-0.65cm}
			\includegraphics[width=1.04 \linewidth]{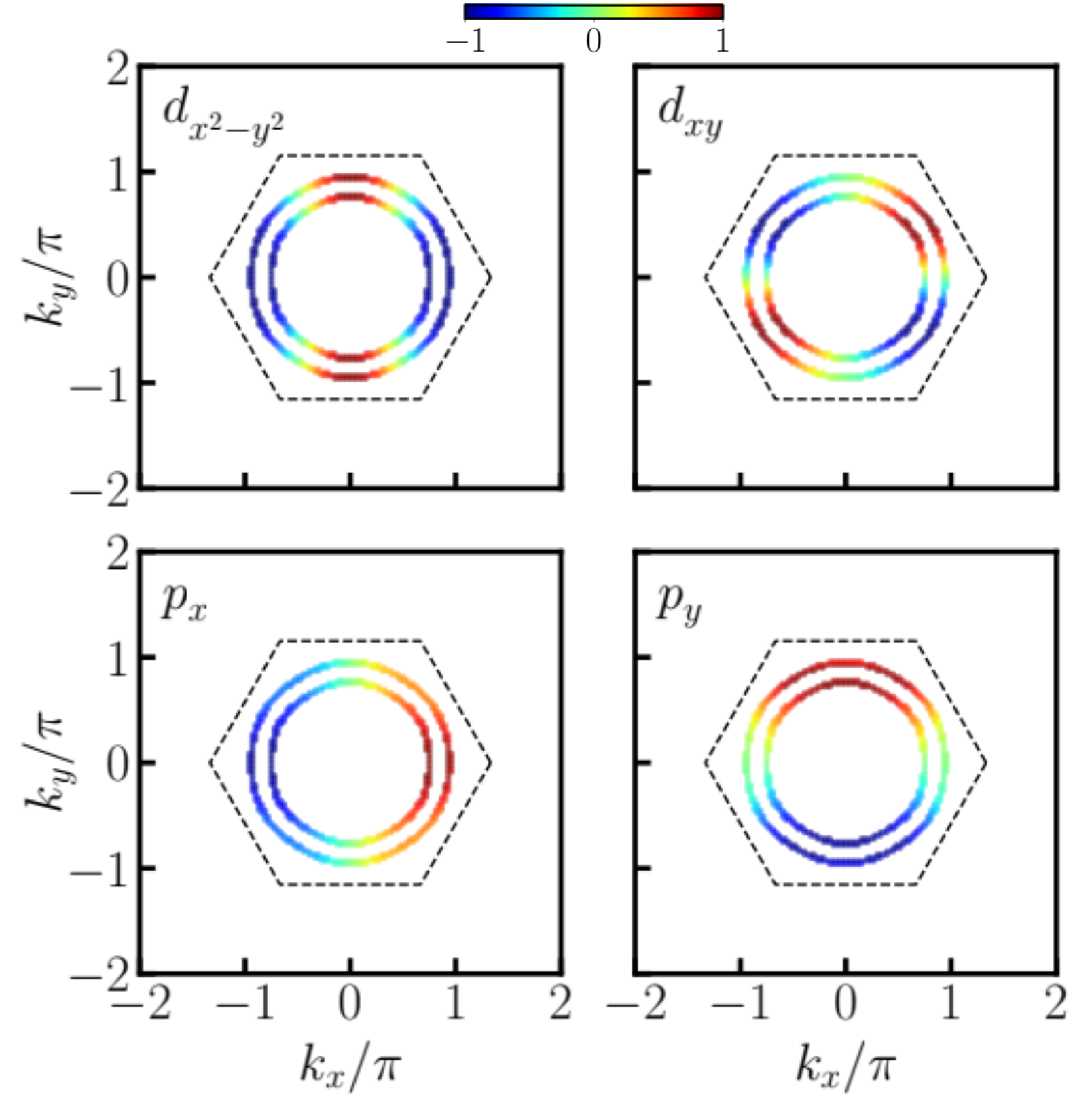} 
		\end{center}
		\caption{ 
			The normalized superconducting gap functions in the singlet (upper row) and triplet (lower row) channels for $t^{}_1=1$, $t^{}_2=0$, and $V^{}_{\rm SOC}=0.3$, at doping level $\delta=0$.
			Rashba SOC violates parity, which results in parity mixing in the superconducting order parameter.
			Due to spontaneously broken time-reversal symmetry, the superconducting gap function in the singlet and triplet channels are $d^{}_{x^2-y^2}\pm {\mi}d^{}_{xy}$ and $p^{}_{x}\pm {\mi}p^{}_{y}$, respectively.
		}
		\vspace{0.73cm}
		\label{Fig:SC_gap_FS_t2_03_SOC_03}
	\end{figure}
	%%%%%%%%%%%%%%%%%%%%%%%%%%%%%%%%%%%%%%%%%%%%%%%%%%%
	%%%%%%%%%%%%%%%%%%%%%%%%%%%%%%%%%%%%%%%%%%%%%%%%%%%
	%
		%
	%%%%%%%%%%%%%%%%%%%%%%%%%%%%%%%%%%%%%%%%%%%%%%%%%%%
	%%%%%%%%%%%%%%%%%%%%====== figure ======%%%%%%%%%%%%%%%%%%%
	%%%%%%%%%%%%%%%%%%%%%%%%%%%%%%%%%%%%%%%%%%%%%%%%%%%
	\begin{figure}[t]
		\begin{center}
			\hspace{-0.65cm}
			\includegraphics[width=1.04 \linewidth]{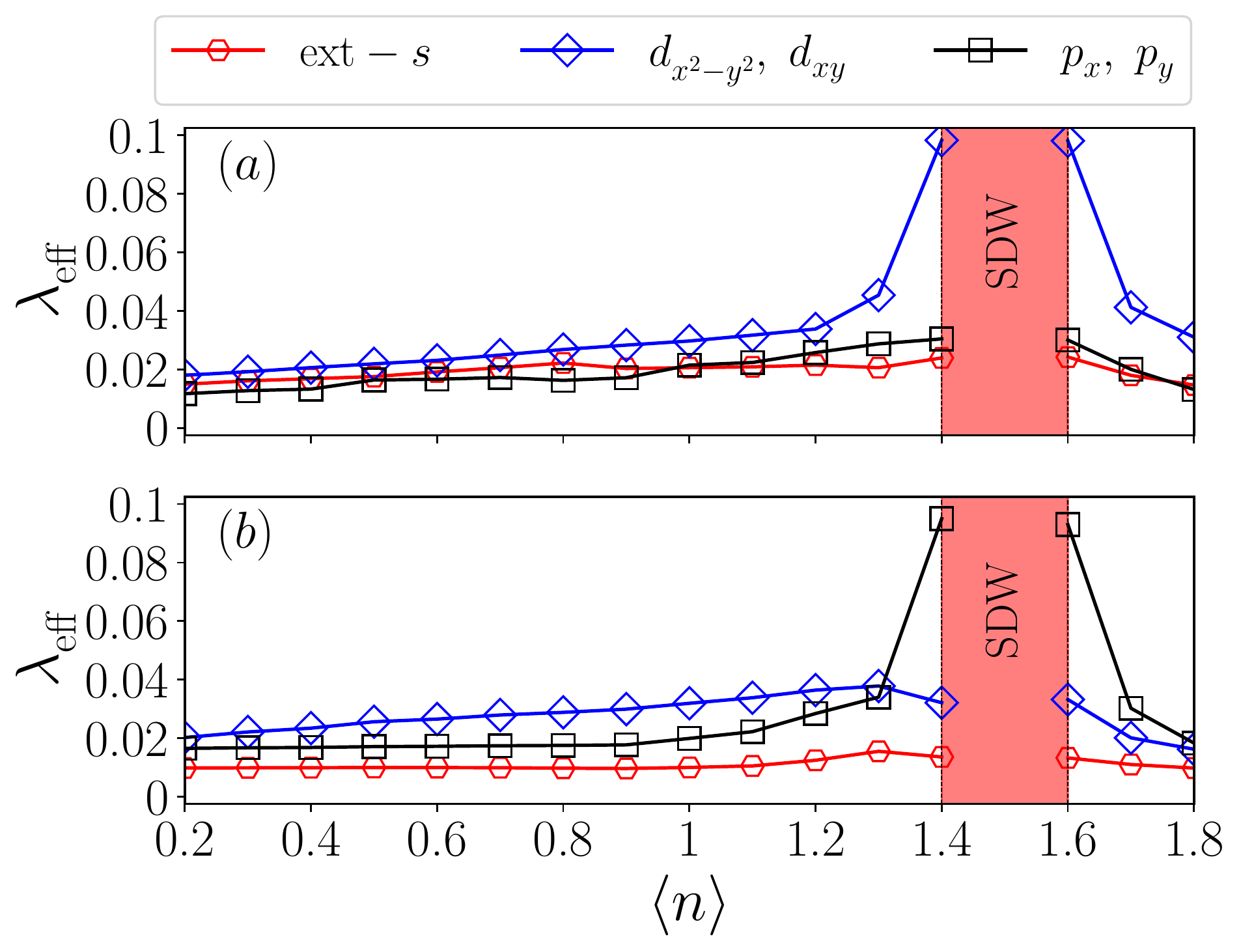} 
		\end{center}
		\caption{
			Magnitude of the eigenvalue \eqref{Eq:SC_Lambda_ss'} for different superconducting channels as a function of doping for (a) $t_2=V_{\rm SOC}=0$ and (b) $t_2=0$, $V_{\rm SOC}=0.3$ with $U=2$. The red region denotes the doping levels where magnetic order competes with superconductivity.
		}
		\vspace{0.73cm}
		\label{Fig:Lambda}
	\end{figure}
	%%%%%%%%%%%%%%%%%%%%%%%%%%%%%%%%%%%%%%%%%%%%%%%%%%%
	%%%%%%%%%%%%%%%%%%%%%%%%%%%%%%%%%%%%%%%%%%%%%%%%%%%
		%
	%%%%%%%%%%%%%%%%%%%%%%%%%%%%%%%%%%%%%%%%%%%%%%%%%%%
	%%%%%%%%%%%%%%%%%%%%====== figure ======%%%%%%%%%%%%%%%%%%%
	%%%%%%%%%%%%%%%%%%%%%%%%%%%%%%%%%%%%%%%%%%%%%%%%%%%
	\begin{figure}[t]
		\begin{center}
			\hspace{-0.65cm}
			\includegraphics[width=1.04 \linewidth]{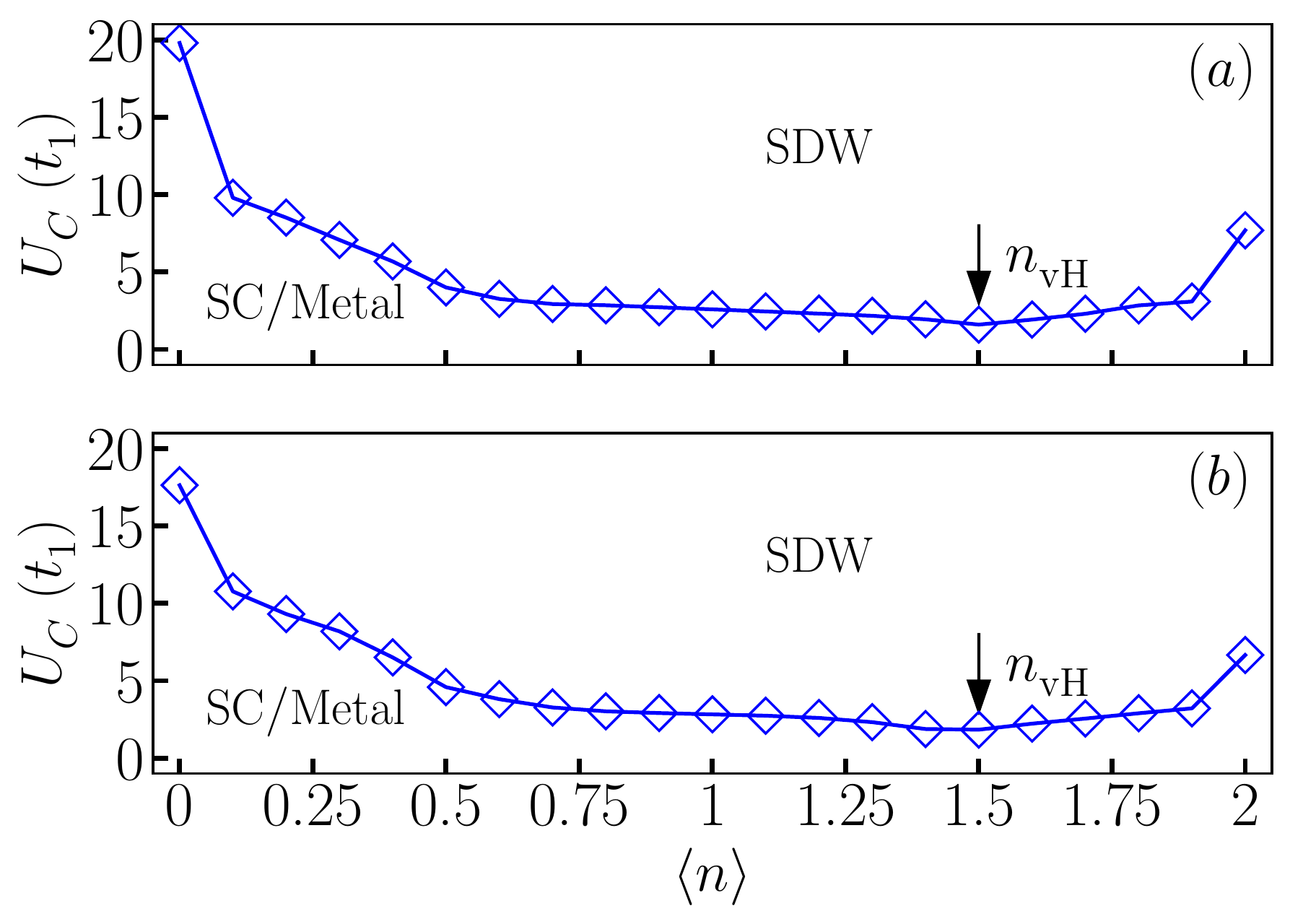} 
		\end{center}
		\caption{
			Critical line separating an SDW phase from a superconducting/metallic phase for (a) $t_2=V_{\rm SOC}=0$ and (b) $t_2=0$, $V_{\rm SOC}=0.3$. The critical line is determined by finding the Coulomb interaction $U_c$ for each doping level at which the RPA spin susceptibility diverges.
		}
		\vspace{0.73cm}
		\label{Fig:Phase_diag}
	\end{figure}
	%%%%%%%%%%%%%%%%%%%%%%%%%%%%%%%%%%%%%%%%%%%%%%%%%%%
	%%%%%%%%%%%%%%%%%%%%%%%%%%%%%%%%%%%%%%%%%%%%%%%%%%%
	%
	%
	\subsection{Results}
	First, we consider the symmetry of the gap function. Fig.~\ref{Fig:SC_gap_FS_t2_03_SOC_0} shows the dominant superconducting pairing for $t^{}_2=0$, $V^{}_{\rm SOC}=0$, and doping $\delta=0$.
	Here, the normalized magnitude of the superconducting gap function on the Fermi surface is depicted.
	Since Rashba coupling is absent, spin is a good quantum number and we find that the leading superconducting instability is in the singlet channel with degenerate $d^{}_{x^2-y^2}$- and $d^{}_{xy}$-wave pairings. 
	This degeneracy originates from the C$^{}_{3v}$ symmetry of the triangular lattice.
	Minimization of the energy gives a chiral $d+id$ pairing, which is a consequence of the geometrical frustration of the triangular lattice~\cite{Lu_PRB_2018}.
	Moreover, we find that this result remains true irrespective of the doping level. 
	If the system picks one of the degenerate states $d^{}_{x^2-y^2}\pm{\mi}d^{}_{xy}$ then TRS will be spontaneously broken. 
	This type of chiral pairing belongs to the $C$ class of topological superconductors, which are characterized by a $\mathbb{Z}$ topological index.
	Next, we consider the influence of the SOC on the superconducting gap function. Since spin is then no longer a good quantum number we expect that singlet and triplet components are mixed in general. Indeed for $t^{}_2=0$, $V^{}_{\rm SOC}=0.3$, and $U=2$ we find both d-wave and p-wave superconductivity, see Fig.~\ref{Fig:SC_gap_FS_t2_03_SOC_03}.
	As in the case without SOC, the pairing symmetry in the singlet channel has the chiral form $d^{}_{x^2-y^2}\pm{\mi}d^{}_{xy}$.
	Since Rashba SOC also breaks parity, we find that the even-parity spin singlet and odd-parity spin-triplet pairings are mixed together.
	Our calculations show that the leading pairing in the triplet channel has both $p^{}_x$ and $p^{}_y$ symmetries at the same eigenvalue $\lambda$.
	In a similar manner to the singlet channel, the minimization of energy in the triplet channel will thus show a chiral texture of the form $p^{}_x+{\mi}p^{}_y$.
	This means that the symmetry of Cooper pairs in the triplet channel belongs to $E^{}_1$ irrep.  
	In contrast to the case of a square lattice, in which the triplet part of the superconducting gap has a helical characteristic~\cite{Rachel_PRB_2020}, the spontaneous breaking of TRS in the superconducting gap function for a triangular lattice generates a chiral triplet order parameter.
	As a result, the precise form of the superconducting gap function is $(d^{}_{x^2-y^2}+{\mi}d^{}_{xy})+(p^{}_x+{\mi}p^{}_y)$-wave.
	Note, however, that our calculations show that for the parameters chosen in Fig.~\ref{Fig:SC_gap_FS_t2_03_SOC_03} the singlet pairing is dominant. 
	
	More generally speaking, the magnitude of the eigenvalues in the singlet and triplet channels will depend on the hopping amplitude $t_2$, the strength of SOC, the Hubbard interaction, and the doping level. 
	As shown in Fig.~\ref{Fig:Lambda}, the d-wave instability is always dominant in the absence of SOC. 
	With SOC, on the other hand, we find that a d-wave instability is dominant in the hole-doped case while a p-wave instability dominates for electron doping close to the SDW instability. 
	This result is consistent with our earlier results for the spin susceptibility: For electron doping levels in between the two VHSs, the triplet  pairing is dominant due to the presence of strong ferromagnetic fluctuations.
	We stress again that in this case the singlet and triplet channels will always be mixed and only the relative level of admixture is changing with doping. Using that $T_c\propto \omega_c \exp(-1/\lambda_{\rm eff})$ we can also obtain a rough estimate for the transition temperature $T_c$ where the cutoff frequency $\omega_c$ is of the order of the bandwidth. If we assume a bandwidth in the range of 1 eV we need a $\lambda_{\rm eff}\geq 0.1$ to get a transition temperature in the range of a few Kelvin. We see in Fig.~\ref{Fig:Lambda} that such values for $\lambda_{\rm eff}$ are indeed reached for the leading superconducting pairings close to the regime where the system shows an SDW instability. 
	
	We can also use the RPA results to determine the ground state phase diagram by finding the critical Hubbard interaction strength $U_c$ at which the spin susceptibility diverges which indicates an instability towards spin-density wave (SDW) order. In this picture there is a regime $U<U_c$ where superconductivity is established---note that we are discussing the ground state phase diagram: the critical temperature $T_c$ might be extremely small and not accessible in experiment which is why we denote this region by SC/metal in Fig.~\ref{Fig:Phase_diag}---and a regime $U>U_c$ where magnetic order sets in. The critical line separating the two phases obtained in this manner is shown in Fig.~\ref{Fig:Phase_diag}.

	In both cases, there is a minimum at a filling which corresponds to the van-Hove singularity in the DOS, see Fig.~\ref{Fig:Geometry_Bands_DOS_FS}.

	\section{Quasiparticle interference}
	\label{QPI}
	In this section we want to discuss an experimental technique which can potentially be used to investigate the superconducting state. We note that quasiparticle interference is an STM technique which does require access to the surface of the bilayer and is therefore not suitable for a heterostructure with both a top and a bottom gate. The results presented in this section are therefore, for now, purely theoretical and applicable for a hypothetical system where the layer realizing the 2d triangular Hubbard model is accessible by STM.
	
	%%
	%%%%%%%%%%%%%%%%%%%%%%%%%%%%%%%%%%%%%%%%%%%%%%%%%%%
	%%%%%%%%%%%%%%%%%%%%====== figure ======%%%%%%%%%%%%%%%%%%%
	%%%%%%%%%%%%%%%%%%%%%%%%%%%%%%%%%%%%%%%%%%%%%%%%%%%
	\begin{figure}[t]
		\begin{center}
			\vspace{0.5 cm}
			\hspace{-0.65cm}
			\includegraphics[width=1.04 \linewidth]{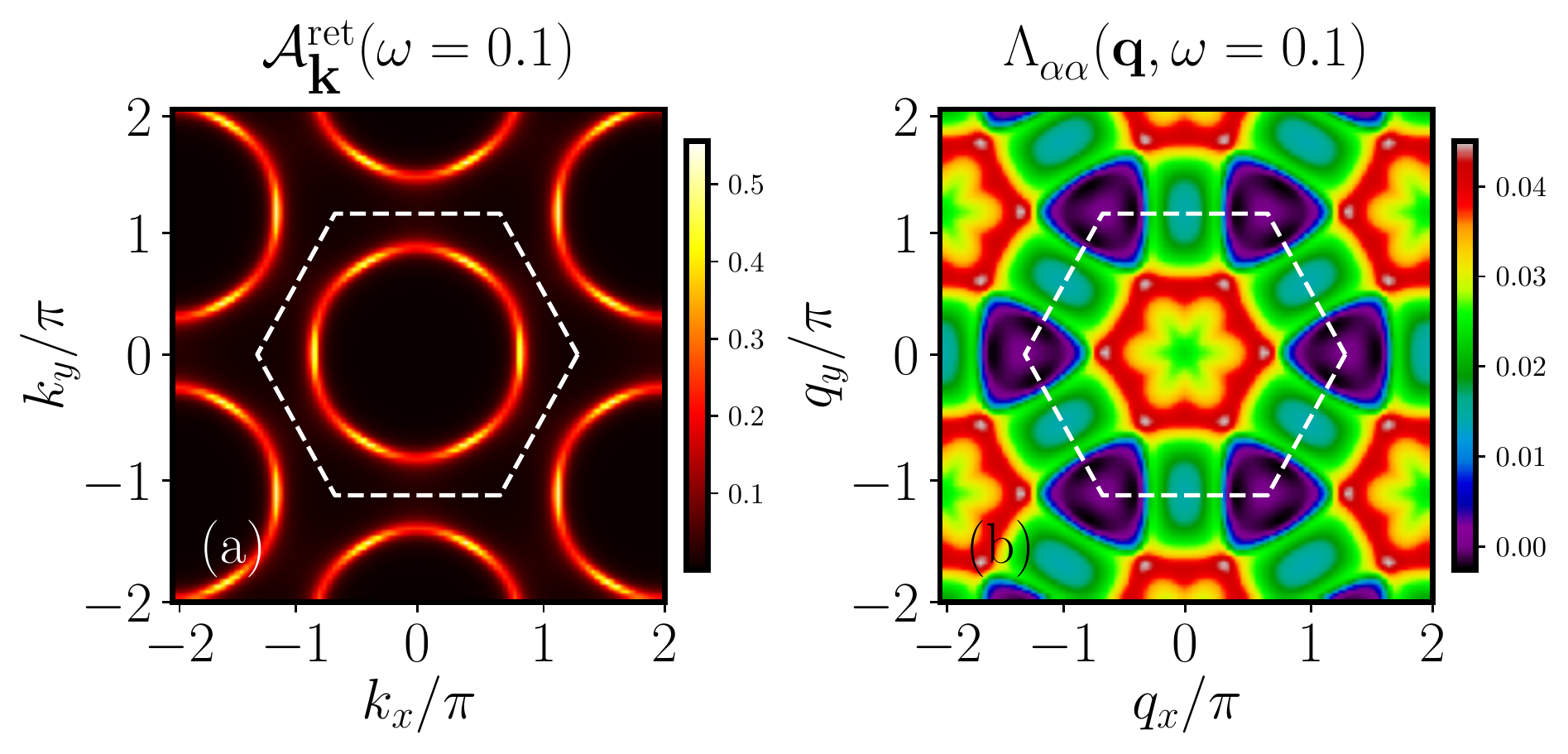} 
		\end{center}
		\vspace{-0.7 cm}
		\caption{ 
			(a) Spectral function of the
			superconducting state in the absence of SOC ($V^{}_{\rm SOC}=0$) for $t^{}_1=1$, $\Delta^{}_{\rm S}=0.2$, and $\Delta^{}_{\rm T}=0$.
			(b) Total QPI for both charge ($\eta=0$), and spin ($\eta=x$, $y$, and $z$) channels at $\omega=0.1$.
			Note that because of preserved SU$(2)$ symmetry, all components of the QPI are the same.
		}
		\vspace{0.73cm}
		\label{Fig:QPI_no_SC}
	\end{figure}
	%%%%%%%%%%%%%%%%%%%%%%%%%%%%%%%%%%%%%%%%%%%%%%%%%%%
	%%%%%%%%%%%%%%%%%%%%%%%%%%%%%%%%%%%%%%%%%%%%%%%%%%%
	%
	
	\subsection{Method}
	First, we briefly review how to calculate the relevant quantities measured experimentally. The STM-based quasiparticle interference (QPI) technique is one of the most powerful methods to discriminate between different gap symmetries in a wide variety of superconducting materials.
	In the presence of magnetic or non-magnetic impurities, the local density of states (LDOS) will be modulated.
	Consequently, the superposition of incident and scattered quasiparticles generate interference patterns, carrying information about the Fermi surface and the superconducting gap function.
	The QPI pattern is obtained from the full Green's function, which is determined by considering the effect of scattering from random charge and spin impurities at the surface.
	The scattering potential is given by~\cite{Mahan_ManyBody_2000}
	%
	%
	%===========================================================
	%========================Equation===========================
	%===========================================================
	\begin{equation}
		{\cal H}^{}_{\rm imp}=
		\sum_{\bk\bq\eta}
		V^{\eta}_{\bq}
		\hat{S}^{}_{\eta}
		\Psi^{\dagger}_{\bq+\bk}
		\hat{\rho}^{}_{\eta}
		\Psi^{}_{\bq},
		\label{Eq:Scattering_Potential}
	\end{equation}
	%===========================================================
	%
	where $\Psi^{\dagger}_{\bk}=(c^{\dagger}_{\bk\uparrow},c^{\dagger}_{\bk\downarrow},c^{}_{-\bk\uparrow},c^{}_{-\bk\downarrow})$ represents the Nambu space describing the electron-hole symmetric basis.
	Within this space, $\hat{\rho}^{}_{\eta=0,x,y,z}$ represent $4\times 4$ matrices with $\lbrace \hat{\rho}^{}_{\eta} \rbrace=( \tau^{}_z\otimes\sigma^{}_0,\tau^{}_0\otimes\sigma^{}_x,\tau^{}_z\otimes\sigma^{}_y,\tau^{}_0\otimes\sigma^{}_z)$, in which $\tau$ and $\sigma$ are Pauli matrices in particle-hole and spin spaces, respectively.
	It should be mentioned that $\tau^{}_0$ and $\sigma^{}_0$ are $2\times 2$ identity matrices. 
	Moreover, $\lbrace S^{}_{\eta} \rbrace=(1,\bS)$ with $\bS$ denoting the spin of the magnetic impurity.
	The potential of a non-magnetic impurity is given by $V^{\eta=0}_{\bq}$, while a magnetic impurity potential is given by an exchange scattering ($V^{\eta\neq 0}_{\bq}=V^{}_{\rm ex}$) from an impurity with spin $\bS$.
	Within the second order Born approximation, the Fourier transform of the modulation of LDOS at frequency $\omega$ is~\cite{Scalapino_PRL_2003}
	%
	%
	%===========================================================
	%========================Equation===========================
	%===========================================================
	\begin{equation}
		dN^{}_{\eta\eta'}(\bq,\omega)=
		-\frac{1}{\pi}
		V^{\eta'}_{\bq}
		{\rm Im}
		\Big[
		\Lambda^{}_{\eta\eta'}(\bq,\omega)
		\Big],
		\label{Eq:LDOS_Modulation}
	\end{equation}
	%===========================================================
	%
	in which
	%
	%===========================================================
	%========================Equation===========================
	%===========================================================
	\begin{equation}
		\Lambda^{}_{\eta\eta'}(\bq,\omega)=
		\frac{1}{N}
		\sum_{\bk}
		{\rm Tr}^{}_{\sigma}
		\Big[
		\hat{\cal P}^{}_{}
		\hat{\rho}^{}_{\eta}
		\hat{G}^{}_{\bk}(\mi\omega)
		\hat{\rho}^{}_{\eta'}
		\hat{G}^{}_{\bk+\bq}(\mi\omega)
		\Big]^{}_{\mi\omega\rightarrow\omega+{\mi}0^+}.
		\label{Eq:QPI_Kernel}
	\end{equation}
	%===========================================================
	%
	In this equation, $\hat{\cal P}=(\tau^{}_0+\tau^{}_z)\otimes\sigma^{}_0/2$ is a projection operator.
	In addition, $\hat{G}^{}_{\bk}(\mi\omega)=[\mi\omega-\hat{\cal H}^{\rm BdG}_{\bk}]^{-1}_{}$ is the Matsubara Green's function of the superconducting state, in which the Boguliubov de-Gennes Hamiltonian is given by
	%
	%
	%===========================================================
	%========================Equation===========================
	%===========================================================
	\begin{equation}
		\hat{\cal H}^{\rm BdG}_{\bk}=
		\begin{bmatrix}
			\hat{\cal H}^{0}_{\bk} & \hat{\Delta}^{}_{\bk}
			\\
			\hat{\Delta}^{\dagger}_{\bk} & -\hat{\cal H}^{0*}_{-\bk}
		\end{bmatrix}.
		\label{Eq:BdG_Hamiltonioan}
	\end{equation}
	%===========================================================
	%
	Here, $\hat{\cal H}^{0}_{\bk}$ represents the non-interacting Hamiltonian of the normal state.
	Moreover, the momentum resolved spectral function is represented by ${\cal A}^{\rm ret}_{\bk}(\omega)=-2{\rm Im}[{\rm Tr}^{}_{\sigma} \hat{G}^{}_{\bk}({\mi}\omega)]^{}_{{\mi \omega}\rightarrow \omega+{\mi}0^{+}}$ and $\hat{\Delta}^{}_{\bk}={\mi}[\psi^{}_{\bk}\hat{\sigma}^{}_0+\bd^{}_{\bk}\cdot\hat{\boldsymbol{\sigma}}]\hat{\sigma}^{}_y$ is the matrix of superconducting gap functions in spin space.
	Here, $\psi^{}_{\bk}$, and $\bd^{}_{\bk}$ represent the singlet and triplet parts of the superconducting gap function with amplitudes $\Delta^{}_{\rm S}$ and $\Delta^{}_{\rm T}$, respectively, whose explicit forms are given by table.~\ref{Tab:SC_gap}.
	Spontaneous breaking of time-reversal symmetry (TRS) in a triangular lattice generates a degenerate superconducting ground state for both singlet and triplet channels.
	The projection of the superconducting gap function then gives rise to
	%
	%
	%===========================================================
	%========================Equation===========================
	%===========================================================
	\begin{equation}
		\Delta^{}_{\bk s}=\psi^{}_{\bk}+s |\bd^{}_{\bk}|,
		\label{Eq:Sc_Gap_Bands}
	\end{equation}
	%===========================================================
	%
	in which $\Delta^{}_{\bk s}$ denotes the magnitude of the gap function on the Fermi surface of band $s$.
	In this picture, the energy dispersion of quasiparticles in band $s$ is given by $E^{}_{\bk s}=\sqrt{\xi^{2}_{\bk s}+\Delta^{2}_{\bk s}}$. 
	%Eq.~(\ref{Eq:SC_dispersion}).
	%
	Using Eq.~(\ref{Eq:QPI_Kernel}), the QPI patterns for charge and spin channels can be shown to be given by~\cite{Akbari_epl_2013}
	%
	%===========================================================
	%========================Equation===========================
	%===========================================================
	\begin{align}
		\begin{aligned}
			\Lambda^{}_{00}(\bq,\omega)
			&=
			\frac{1}{N}\sum_{\bk ss'}
			[
			1+ss'
			(
			\hat{\bg}^{}_{\bk}
			\cdot
			\hat{\bg}^{}_{\bk+\bq}
			)
			]
			{\cal K}^{\bk\bq}_{ss'}(\omega),
			\\
			\Lambda^{}_{uu}(\bq,\omega)
			&=
			\frac{1}{N}\sum_{\bk ss'}
			[
			1-ss'
			(
			\hat{\bg}^{}_{\bk}
			\cdot
			\hat{\bg}^{}_{\bk+\bq}
			-2
			\hat{\bg}^{u}_{\bk}
			\hat{\bg}^{u}_{\bk+\bq}
			)
			]
			{\cal K}^{\bk\bq}_{ss'}(\omega),
		\end{aligned}
		\label{Eq:Lambda_ii}
	\end{align}
	%===========================================================
	%
	where the index $u=(x,y,z)$ is running over the spatial components of the spin.
	In addition, the QPI kernel function ${\cal K}^{\bk\bq}_{ss'}(\omega)$ is given by
	%
	%
	%===========================================================
	%========================Equation===========================
	%===========================================================
	\begin{equation}
		{\cal K}^{\bk\bq}_{ss'}(\omega)=
		\frac{
			(\omega+\xi^{}_{\bk s})
			(\omega+\xi^{}_{\bk+\bq s'})
			-
			\Delta^{}_{\bk s}
			\Delta^{}_{\bk+\bq s'}
		}
		{
			(\omega^2-E^{2}_{\bk s})
			(\omega^2-E^{2}_{\bk+\bq s'})
		}.
		\label{Eq:QPI_Kernel_expression}
	\end{equation}
	%===========================================================
	%
		%	%%
	%%%%%%%%%%%%%%%%%%%%%%%%%%%%%%%%%%%%%%%%%%%%%%%%%%%
	%%%%%%%%%%%%%%%%%%%%====== figure ======%%%%%%%%%%%%%%%%%%%
	%%%%%%%%%%%%%%%%%%%%%%%%%%%%%%%%%%%%%%%%%%%%%%%%%%%
	\begin{figure*}[t]
		\begin{center}
			\hspace{-0.8cm}
			\includegraphics[width=1.05 \linewidth]{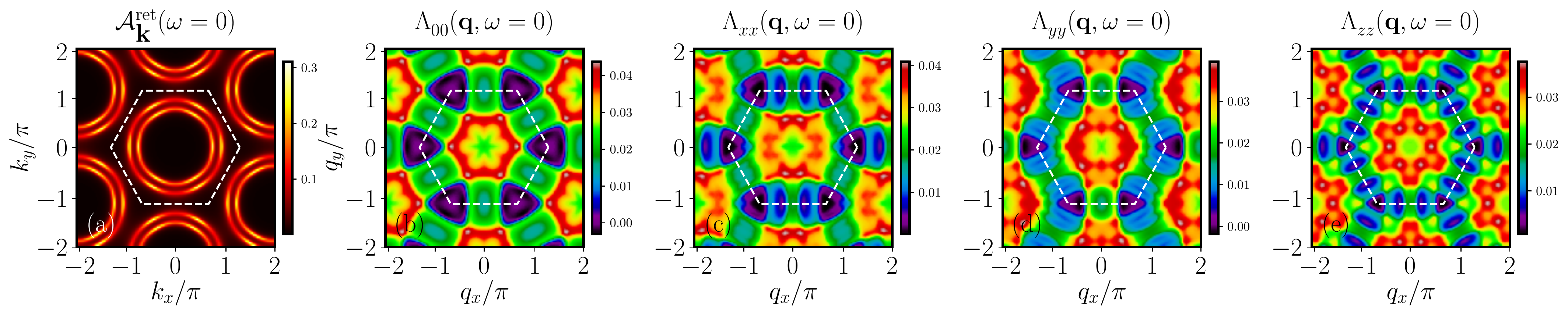} 
		\end{center}
		\vspace{-0.7 cm}
		\caption{ 
			(a) Spectral function of the superconducting state in the presence of SOC ($V^{}_{\rm SOC}=0.3$) for $t^{}_1=1$, $\Delta^{}_{\rm S}=0.2$, and $\Delta^{}_{\rm T}=0.05$ at $\omega=0.1$.
			The areas with high spectral weight dominate the quasiparticle scattering events.
			(b) Total QPI pattern $\Lambda^{}_{00}(\bq,\omega)$ in the charge channel carrying the crystal symmetry. 
			(c,d) QPI pattern in the spin channel due to the scattering by a magnetic impurity with spins $S^{}_x$ and $S^{}_y$, respectively.
			Since SOC breaks the SU$(2)$ symmetry,  $\Lambda^{}_{xx}(\bq,\omega)$ and $\Lambda^{}_{yy}(\bq,\omega)$ do not follow the crystal point group symmetry.
			(e) The interference pattern $\Lambda^{}_{zz}(\bq,\omega)$ corresponding to a magnetic impurity with spin $S^{}_z$  shows again the crystal symmetry.
		}
		\vspace{0.73cm}
		\label{Fig:QPI_with_SC}
	\end{figure*}
	%%%%%%%%%%%%%%%%%%%%%%%%%%%%%%%%%%%%%%%%%%%%%%%%%%%
	%%%%%%%%%%%%%%%%%%%%%%%%%%%%%%%%%%%%%%%%%%%%%%%%%%%
	%
	
	\subsection{Results}
	%
	% At this step, we provide an experimental method to detect the properties of superconducting gap function in our effectively triangular lattice.
	% %
	% The STM-based QPI is one of the outstanding approaches
	% used to study the possible existence of a superconducting
	% state and investigate its pairing symmetry.
	% %
	% It can provide many useful
	% information about the structure of gap symmetry and its
	% dependence on the phase of the superconducting order parameter.
	%
	In Fig.~\ref{Fig:QPI_no_SC}(a), an intensity plot for the superconducting spectral function at $\omega=0.1$ for $t^{}_2=0$, $V^{}_{\rm SOC}=0$, $\Delta^{}_{\rm S}=0.2$, and $\Delta^{}_{\rm T}=0$ is shown.
	In this case, the gap function has the chiral $d^{}_{x^2-y^2}+{\mi}d^{}_{xy}$-wave symmetry.
	The spectral function shows that the areas around the corners of the hexagon have the largest DOS.
	Therefore, the scattering wave-vectors connecting these points are playing the most important role in determining the QPI pattern.
	Because of the preserved SU$(2)$ symmetry, all different components of QPI, including scattering from non-magnetic and magnetic impurities, have the same shapes.
	Fig.~\ref{Fig:QPI_no_SC}(b) shows the QPI pattern $\Lambda^{}_{\eta\eta}(\bq,\omega=0.1)$ ($\eta=\lbrace 0,x,y,z\rbrace$).
	The pattern is consistent with the point-group symmetry of the lattice.
	Moreover, since the superconducting gap function has a fully gapped structure, the QPI pattern resembles the shape of the Fermi surface, including some dominant peaks at the corners of the hexagon.

	Finally, in Figs.~\ref{Fig:QPI_with_SC}, the superconducting spectral function and the QPI patterns generated by non-magnetic and magnetic impurities in the presence of SOC are reported.
	Fig.~\ref{Fig:QPI_with_SC}(a) shows the spectral function for  $t^{}_2=0$, $V^{}_{\rm SOC}=0.3$, $\Delta^{}_{\rm S}=0.2$, and $\Delta^{}_{\rm T}=0.05$ at frequency $\omega=0.1$.
	In this situation, the Cooper pairs have the symmetry of a $(d^{}_{x^2-y^2}+{\mi}d^{}_{xy})+(p^{}_x+{\mi}p^{}_y)$-wave.
	This plot carries the information about the Fermi surface together with information about the spectral weights.
	The QPI pattern for the scattering of the superconducting quasiparticles from a non-magnetic impurity is shown in Fig.~\ref{Fig:QPI_with_SC}(b).
	The dominant sharp peaks originate from intra-band scattering.
	However, there are also some sub-leading peaks around the $\boldsymbol{\Gamma}$ point, which can be attributed to inter-band scattering.
	Figs.~\ref{Fig:QPI_with_SC}(c,d) show $\Lambda^{}_{xx}(\bq,\omega=0.1)$, and $\Lambda^{}_{yy}(\bq,\omega=0.1)$, respectively.
	We observe that these two QPI patterns have more complicated forms than $\Lambda^{}_{00}(\bq,\omega=0.1)$.
	Eq.~(\ref{Eq:Lambda_ii}) analytically justifies this complex form.
	Since the Rashba SOC only has in-plane components, the QPI patterns generated by magnetic impurities with spins $S^{}_{x}$ and $S^{}_{y}$ will not obey the symmetry of the crystal point-group.
	Lastly, Fig.~\ref{Fig:QPI_with_SC}(e) depicts $\Lambda^{}_{zz}(\bq,\omega=0.1)$.
	Based on Eq.~(\ref{Eq:Lambda_ii}) it is clear that this QPI pattern---resulting from the scattering of superconducting quasiparticles by a magnetic impurity of spin $S^{}_z$---shows again the D$^{}_3$ symmetry of the system.
	%

	%%%%%%%%%%%%%%%%%%%%%%%%%%%%%%%%%%%%%%%%%%%%%%%%%%%%%%%%%%
	\section{Summary and Conclusions}
	We have studied a Rashba-Hubbard Hamiltonian on a triangular lattice as a model for a twisted bilayer of TMD which has a maximum LDOS at ${\cal R}^M_M$ high-symmetry positions.
	In our model, the Rashba SOC is a consequence of depositing the 2D thin layers on a thick substrate.

	In the absence of second neighbor hopping and Rashba SOC, we find incommensurate magnetic fluctuations consistent with earlier results in the literature.
	We then observe that increasing the amplitude of the next-nearest neighbor hopping leads to a reduction of the incommensurate spin fluctuations around the $\bK$ point while amplifying those around the $\bM$ position.

	Looking at electron pairings mediated by these spin fluctuations, we find that in the case without SOC a chiral spin singlet pairing with $d^{}_{x^2-y^2}+{\mi}d^{}_{xy}$ symmetry dominates.
	Upon introducing a Rashba SOC, dominant ferromagnetic fluctuations in the longitudinal susceptibility can arise which are highly sensitive to the level of filling. 
	These fluctuations appear when both hole- and electron-like Fermi surfaces exist.
	In the case with SOC, the superconducting fluctuations become predominantly of 
	$(d^{}_{x^2-y^2}+{\mi}d^{}_{xy})+(p^{}_x+{\mi}p^{}_y)$ character.
	A superconducting phase with this symmetry belongs to the $C$ class of topological superconductors which can be characterized by a $\mathbb{Z}$ topological invariant.
	%
% 	In this manner, our system behaves as a topological quantum Hall state.
	%
	
	In order to experimentally detect gap structures in superconducting materials, the STM-based quasiparticle interference is a powerful tool. We have therefore also examined the QPI response both from non-magnetic and magnetic impurities. The obtained results might serve as a guide for future experiments searching for exotic superconducting phases in tunable bilayer materials. 
	%
	%
	
	%%%%%%%%%%%%%%%%%%%%%%%%%%%%%%%%%%%%%%%%%%%%%%%%%%%%%%%%%%%
	%%%%%%%%%%%%%%%%%%%%%%%%%%%%%%%%%%%%%%%%%%%%%%%%%%%%%%%%%%%
	%%%%%%%%%%%%%%%%%%%%%%%%%%%%%%%%%%%%%%%%%%%%%%%%%%%%%%%%%%%
	%
	\section*{Acknowledgments}
	We are grateful to A.~Greco, Y.~Yanase, A.~Akbari, S.~A.~Jafari, A.~Romer, M.~Malakhov, and M.~N.~Najafi for fruitful 	discussions.  
	J.~S. acknowledges support by the Natural
	Sciences and Engineering Research Council (NSERC, Canada) and by the
	Deutsche Forschungsgemeinschaft (DFG) via Research Unit FOR 2316.
	%

	% 
	%%%%%%%%%%%%%%%%%%%%%%%%%%%%%%%%%%%%%%%%%%%%%%%%%%%%%%%%%%%
	%%%%%%%%%%%%%%%%%%%%%%%%%%%%%%%%%%%%%%%%%%%%%%%%%%%%%%%%%%%
	\bibliography{Refs}
	%%%%%%%%%%%%%%%%%%%%%%%%%%%%%%%%%%%%%%
	%%%%%%%%%%%%%%%%%%%%%%%%%%%%%%%%%%%%%%
	
\end{document}